\documentclass[journal]{IEEEtran}
\usepackage{nomencl}
\makenomenclature

\usepackage[utf8]{inputenc}
\usepackage{algorithm}% http://ctan.org/pkg/algorithms
\usepackage{algpseudocode}% http://ctan.org/pkg/algorithmicx
\usepackage{xcolor}% http://ctan.org/pkg/xcolor
\usepackage{listings}
\usepackage[noadjust]{cite}
\usepackage[normalem]{ulem}
\usepackage{paralist}
\usepackage{mathtools}
\usepackage{amsthm}
\usepackage{bm}
\usepackage{physics}
\usepackage{textcomp}
\usepackage{subcaption}
\usepackage{atbegshi}
\usepackage{amsfonts}
\usepackage{multirow}
\usepackage{soul}
\usepackage{hhline}  % added by Jarek

\newcommand{\df}{\coloneqq}

\renewcommand{\vec}[1]{\mathbf{#1}}

\ifCLASSINFOpdf
\else
\fi
\AtBeginDocument{\AtBeginShipoutNext{\AtBeginShipoutDiscard}}

\begin{document}
	\title{Revisiting the nonlinear Gaussian noise model: The case of hybrid fiber spans}
	
	\author{I.~Roudas,~\IEEEmembership{Member,~IEEE}, J.~Kwapisz, and X.~Jiang,~\IEEEmembership{Member,~IEEE}}% <-this % stops a space
	\thanks{I.~Roudas is with the Department of Electrical and Computer Engineering, Montana State University, Bozeman, MT 59717, USA email: ioannis.roudas@montana.edu}
\thanks{J.~Kwapisz is with the 	Department of Mathematical Sciences, Montana State University, Bozeman, MT 59717, USA email: jarek@math.montana.edu}
\thanks{X.~Jiang is with the Department of Engineering and Environmental Science, College of Staten Island, City University of New York, Staten Island, NY 10314, USA email: jessica.jiang@csi.cuny.edu}


	\markboth{JOURNAL OF LIGHTWAVE TECHNOLOGY, \today} %February~2018}%
	{\MakeLowercase{\textit{I. Roudas et al.}}: Performance of Coherent Optical Communication Systems With Hybrid Fiber Spans}
	



	\maketitle
	
	\begin{abstract}
We rederive from first principles and generalize the theoretical framework of the nonlinear Gaussian noise model to the  case of coherent optical systems with multiple fiber types per span and ideal Nyquist spectra. We focus on the accurate numerical evaluation of the  integral for the nonlinear noise variance for hybrid fiber spans. This task consists in addressing four computational aspects: 
\begin{inparaenum}[(i)]
	\item Adopting a novel transformation of variables (other than using hyperbolic coordinates) that changes the integrand to a more appropriate form for numerical quadrature; 
	\item Evaluating analytically the integral at its lower limit, where the integrand presents a singularity; 
	\item Dividing the interval of integration into subintervals of size $\pi$ and approximating the integral in each subinterval by using various algorithms; and 
	\item Deriving an upper bound  for the relative error when the interval of integration is truncated in order to accelerate computation.
\end{inparaenum}

We apply the proposed model to coherent optical communications systems with hybrid fiber spans composed of quasi-single-mode fiber and single-mode fiber segments. The accuracy of the final analytical relationship for the nonlinear noise variance in long-haul coherent optical communications systems with hybrid fiber spans is checked using the split-step Fourier method and Monte Carlo simulation. It is shown to be adequate to within 0.1 dBQ for the determination of the optimal fiber segment lengths per span that maximize system performance. 	
\end{abstract}
	
%	
%	
%	
%	
%	
%	
%	% For peer review papers, you can put extra information on the cover
%	% page as needed:
%	% \ifCLASSOPTIONpeerreview
%	% \begin{center} \bfseries EDICS Category: 3-BBND \end{center}
%	% \fi
%	%
%	% For peerreview papers, this IEEEtran command inserts a page break and
%	% creates the second title. It will be ignored for other modes.
%	\IEEEpeerreviewmaketitle
%	

%\newline
%\vspace{2 cm}

\section{Introduction}
\IEEEPARstart{O}{ne}  of the most important theoretical achievements of recent years in optical telecommunications was the approximate solution of the nonlinear Schr\"{o}dinger equation \cite{agrawalNL} and its vector counterpart, the Manakov equation \cite{Wai_JLT_96}. More specifically, many alternative analytical formalisms, e.g., \cite{Chen:OpEx10}, \cite{Shieh_PJ_11}, \cite{Poggiolini_PTL_11}, \cite{Carena_JLT_12}, \cite{Poggiolini:12}, \cite{johannisson2013perturbation}, \cite{Poggiolini:14},  \cite{Mecozzi_JLT_12}, \cite{Dar:13}, \cite{Carena:14}, \cite{serena2015time}, \cite{Ghazisaeidi:17}, have been proposed for the estimation of the impact of distortion due to Kerr nonlinearity on the performance of coherent optical communications systems with no inline dispersion compensation. Among those, the nonlinear Gaussian noise model (see review papers \cite{Poggiolini:12}, \cite{Poggiolini:14}), was  established in the consciousness of the scientific community as an industry standard, due to its relative simplicity compared to other, more sophisticated but more accurate, models, e.g., \cite{Mecozzi_JLT_12}, \cite{Dar:13}. 

The nonlinear Gaussian noise model was originally developed for a single fiber type per span, lumped optical amplifiers, and ideal Nyquist spectra \cite{Poggiolini_PTL_11}. Over the years, it has been constantly revised and has been applied to a variety of system and link configurations, e.g., see \cite{Carena_JLT_12}, \cite{Poggiolini:12}, \cite{Curri:13}, \cite{Poggiolini:14}, \cite{Carena:14}, \cite{Poggiolini:17}, \cite{semrau2018gaussian}, \cite{rabbani2019general}, \cite{poggiolini2019closedform}, \cite{zefreh2019gnmodel}, \cite{zefreh2019closedform}. 

\begin{figure*}[t]
	\centering
	\includegraphics[width=0.9\textwidth]{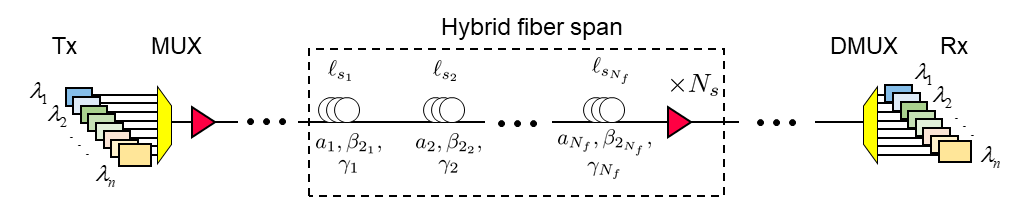}
	\caption{Representative long-haul coherent optical communications system with hybrid fiber spans. }
	\label{fig:BlockDiagram}
\end{figure*}

The nonlinear Gaussian noise-model reference formula (GNRF) \cite{Poggiolini:14} that provides the power spectral density (psd) of nonlinear noise at the end of the link is general enough to encompass the case of hybrid fiber spans, i.e., fiber spans composed of  multiple  segments of different fiber types (Fig. \ref{fig:BlockDiagram}). However, to the best of our knowledge, the application of the nonlinear Gaussian noise-model to coherent optical communications systems with hybrid fiber spans has hardly received any attention to date. Notable exceptions are the following papers: First, Shieh and Chen \cite{Shieh_PJ_11} studied  coherent optical systems with fiber spans consisting of a transmission fiber and a dispersion compensation fiber (DCF). Later on, the papers by Downie et al. \cite{Downie:17} and Miranda et al. \cite{Miranda} were dedicated to hybrid spans comprised of quasi-single-mode  and single-mode fiber segments. More recent publications by Al-Khateeb et al. \cite {Al-Khateeb:18} and Krzczanowicz et al.  \cite{Krzczanowicz:19} focused on hybrid spans for optical phase conjugation  \cite{Al-Khateeb:18} and discrete Raman  amplification  \cite{Krzczanowicz:19}, respectively. 

The aforementioned publications gave diverse expressions for the nonlinear noise coefficient $\tilde{\gamma }$ used to calculate the nonlinear noise variance $\sigma_{NL}^2=\tilde{\gamma }P^3$, where $P$ denotes the total average launch power per channel (in both polarizations). Obviously, these formulas for $\tilde{\gamma }$ are interrelated and their apparent dissimilarities are due to the fact that each individual research group studied a different system topology. Their dissimilarities can be also attributed to the use of  two slightly different formalisms by different authors, i.e.,   \cite{Chen:OpEx10},  \cite{Shieh_PJ_11} and \cite{Poggiolini:12}, \cite{Poggiolini:14}, respectively. Since, on most occasions, only final equations for $\tilde{\gamma }$  were provided without any detailed analytical proof, their direct comparison is difficult. 

%Moreover, methods for the accurate numerical evaluation of the integral for the nonlinear noise variance of coherent optical systems with hybrid fiber spans are not discussed.

Another issue is that numerical quadrature algorithms for the accurate evaluation of the highly oscillatory integral  for the  nonlinear noise coefficient $\tilde{\gamma }$ were not discussed in any of the above papers. One reason that no  special attention has  been devoted to the intricacies of this calculation must be attributed to the fact that a 1-dB error in the nonlinear noise coefficient $\tilde{\gamma }$ results in only 1/3-dB error on optimum effective OSNR \cite{Poggiolini:12}. To the best of our knowledge, only Bononi et al. \cite {Bononi:13} considered in detail the numerical evaluation of the GNRF formula for the case of a single fiber type per span and ideal Nyquist WDM signals.

On a related subject, Poggiolini \cite{Poggiolini:12} recommended to truncate the integration region to reduce the computation time of the foregoing numerical quadrature. Since four-wave mixing (FWM) efficiency quickly drops for increasing values of the mixing frequencies $f_1, f_2$, it was suggested that one could neglect the integration region beyond where the FWM efficiency dropped below a specified level. However, this issue was not investigated thoroughly in \cite{Poggiolini:12} or in subsequent publications.

This paper is intended to fill the aforementioned gaps in the prior literature. First, to reconcile dissimilar formulas derived before for the nonlinear noise variance of coherent optical communications systems with hybrid fiber spans \cite{Shieh_PJ_11}, \cite{Downie:17}, \cite{Miranda},  \cite{Al-Khateeb:18}, \cite{Krzczanowicz:19}, we review and rederive from first principles the theoretical framework of the nonlinear Gaussian noise model for hybrid fiber spans. We find a general expression for the  nonlinear noise variance for the case of an arbitrary number of fiber types per span. Then, we elaborate on the accurate numerical evaluation of the  integral for the nonlinear noise coefficient $\tilde{\gamma }$. The latter task consists in addressing four computational aspects: 
\begin{inparaenum}[(i)]
	\item adopting a novel transformation of variables (other than using hyperbolic coordinates \cite{Poggiolini:12}) that changes the integrand to a more appropriate form for numerical quadrature; 
	\item evaluating analytically the integral at its lower limit, where the integrand presents a singularity;
	\item dividing the interval of integration into panels of size $\pi$ and approximating the integral in each panel by using various algorithms;  and 
	\item deriving an upper bound  for the relative error due to the truncation of the range of integration to accelerate computation.
\end{inparaenum}		

We apply the proposed model to coherent optical communications systems with fiber spans composed of quasi-single-mode fiber and single-mode fiber segments.  The accuracy of the final analytical relationship for the nonlinear noise coefficient in long-haul coherent optical communications systems with hybrid fiber spans is checked using the split-step Fourier method and Monte Carlo simulation. It is shown to be adequate to within 0.1 dBQ for the determination of the optimal fiber segment lengths per span that maximize system performance.

\section{Nonlinear Gaussian noise (GN) model for hybrid fiber spans\label{sec:analytical_model}}

%\noindent 
%\subsection{Problem definition and mathematical notation}
 \subsection{System topology}

\noindent Fig. \ref{fig:BlockDiagram} depicts the block diagram of a representative long-haul coherent optical communication system with hybrid fiber spans. The transmission link of total length $L$ is composed of a concatenation of $N_{s} $ identical spans. Each span has length $\ell _{s}$ and comprises $N_f$ fiber types. Each fiber type is characterized by its nonlinear fiber coefficient $\gamma $, which is a function of the effective mode area $A_{\text{eff}} $ and the nonlinear index coefficient $n_{2} $, its  group velocity dispersion (GVD) parameter ${\beta_2}$ (or, equivalently, its chromatic dispersion parameter  $D$), and its attenuation coefficient $a.$  In what follows, the index $k$ stands for the $k-$th fiber segment per span. For instance, the optical fiber lengths of the $N_f$ segments are $\ell _{s_{1} } ,\ell _{s_{2} } ,\ldots,$  and their effective mode areas are $A_{\text{eff}_1}$, $A_{\text{eff}_2},\ldots$, respectively. The optical fiber is followed by an optical amplifier of gain equal to the span loss $G=\exp\left({\sum_{i=1}^{N_f}{a_i\ell _{s_i}}}\right)$ and noise figure $F_{A} $.

We consider wavelength division multiplexing (WDM) and polarization division multiplexing (PDM) based on ideal Nyquist channel spectra. The latter are created using square-root raised cosine filters with zero roll-off factor at the transmitter and the receiver. Furthermore, we assume that the WDM signal is a superposition of an odd number $N_{ch}$ wavelength channels  with spacing $\Delta \nu=R_{s} .$ We denote by $P$  the total average launch power per channel (in both polarizations) and by $R_{s} $  the symbol rate. We want to evaluate the performance of the center WDM channel at wavelength $\lambda .$

The performance of coherent  optical systems without in-line chromatic dispersion compensation is related to the \textit{effective} optical signal-to-noise ratio (${\rm OSNR}_{\text{eff}}$) at the receiver input. This quantity takes into account the amplified spontaneous emission (ASE)  noise, the multipath interference (MPI) crosstalk (in the case of quasi-single-mode fibers), and the nonlinear distortion. All the above effects can be modeled as independent, zero-mean, complex Gaussian noises with a good degree of accuracy. More specifically, the ${\rm OSNR}_{\text{eff}} $ at  a resolution bandwidth $\Delta \nu _{\text{res}} $ can be well described by the analytical relationship \cite{Mlejnek}
\begin{equation}  
{\rm OSNR}_{\text{eff}} =\frac{P}{\tilde{a}+\tilde{\beta }P+\tilde{\gamma }P^{3} },  
\label{eq:OSNRvsP}
\end{equation} 
where  $\tilde{a}$ is the ASE noise variance, $\tilde{\beta }P$ is the crosstalk variance, and $\tilde{\gamma }P^{3}$ is the nonlinear noise variance. The coefficients $\tilde{a},\tilde{\beta },\tilde{\gamma }$ depend on the fiber and system parameters \cite{Mlejnek}. 

\subsection{Model overview\label{sec:Model_overview}}

\begin{figure*}[t]\centering
	\includegraphics[width=\textwidth]{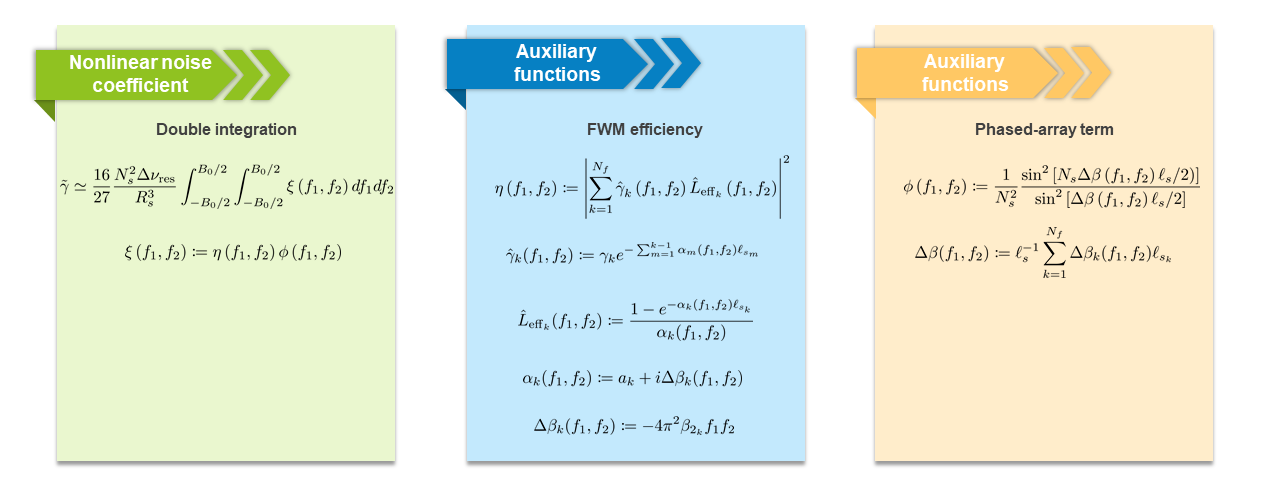}
	\caption{Nonlinear Gaussian noise model for coherent optical communications systems with hybrid fiber spans in a nutshell.}
	\label{fig:SimulationFlowChart}
\end{figure*}

\noindent The purpose of this section is to derive an analytical formula for the nonlinear noise coefficient $\tilde{\gamma }$  in long-haul coherent optical communications systems with hybrid fiber spans.
%\footnote{The term "hybrid" refers  to a fiber span composed of  multiple  segments of different fiber types.  }.

To this end, we will extend the conventional nonlinear Gaussian noise model  \cite{Carena_JLT_12},  \cite{Poggiolini:12}, \cite{Poggiolini:14}, \cite{Poggiolini:17}---initially formulated for a single fiber type per span---to multiple fiber types per span. 

There are three steps used to calculate the variance of nonlinear noise in long-haul coherent optical communications systems with hybrid fiber spans:

\begin{itemize}
	\item Solve the Manakov equation \eqref{eq:Manakov_equation} by using perturbation theory \cite{nayfeh}, \cite{bender2013advanced}, assuming that the fiber attributes, i.e., the attenuation coefficient $a$, the group velocity dispersion (GVD) parameter $\beta_2$, and the nonlinear fiber coefficient $\gamma$, are piecewise constant functions of distance.

	\item Find an analytical expression for the first-order perturbation correction \eqref{eq:c_n1_single_span} to the unperturbed wavefunction.
	
	\item Derive new expressions for the four-wave mixing efficiency term and the phased-array factor and substitute them into the GNRF (see \cite[eq. (18)]{Carena_JLT_12} or  \cite[eq. (1)]{Poggiolini:14}).
	
\end{itemize}

  The final analytical expressions for calculating the nonlinear noise power spectral density in long-haul coherent optical communications systems with hybrid fiber spans are summarized in Fig. \ref{fig:SimulationFlowChart}. We can make the following observations:

 \begin{itemize}
 	
 	\item The sum of contributions from various combinations of three spectral components to the nonlinear noise generated in the middle of the spectrum can be expressed as a double integral (see also \eqref{eq:GNLI}, \eqref{eq:gamma_tilde_ideal_Nyquist} in the main text). 
 	
 	\item The four-wave mixing efficiency term depends on the characteristics of the different fibers per span. It is expressed as a simple
 	formula \eqref{eq:complex_FWM_efficiency_Nf} that depends on two general parameters for each fiber, the complex nonlinear coefficient \eqref{eq:complex_nonlinear_fiber_coefficients_recursion} and the normalized complex effective length \eqref{eq:normalized_complex_effective_lengths_Nf}.
 	
 	\item  The coherent addition of the contributions of successive fiber spans to the total nonlinear noise leads to a phased-array factor (see \eqref{eq:coherent_addition_equal_length_spans}) as in the original nonlinear Gaussian noise model. The only difference is that, in the hybrid fiber span case, the multiple-slit interference term depends on the average phase mismatch \eqref{eq:average_phase_mismatch} of all optical fibers per span.
 \end{itemize}

 	For computational convenience, the double integral can be converted into a single integral by using a transformation of integration variables.  The final integral (see \eqref{eq:nonlinear_noise_coefficient_single_integral} in the main text) is an improper integral of the second kind (i.e., the integrand becomes infinite at the lower end of the integration interval). In addition, the integrand oscillates in the integration interval.  The pseudocode for the numerical quadrature  algorithm is shown below (see Algorithm \ref{alg:gamma}). To accurately compute the integral \eqref{eq:nonlinear_noise_coefficient_single_integral} using numerical quadrature, it is necessary to analytically calculate the contribution in the vicinity of the singularity, then divide the integration interval into $\pi$-subintervals, use a  numerical quadrature method for each subinterval, and add up the results.

%\subsection{Nonlinear Gaussian noise model in a nutshell}
%
% \noindent We want to evaluate the non-linear distortion of a wavelength division multiplexed (WDM) and polarization division multiplexed (PDM)  signal as it propagates through many different fiber-optic segments per span. \hl{We assume that polarization mode dispersion (PMD) and polarization-dependent loss (PDL) are negligible.}

 %\subsection{Motivation}

 \begin{algorithm}%[H]
 	\caption{Pseudocode for the nonlinear noise coefficient.}
 	\label{alg:gamma}	
 	\renewcommand{\thealgorithm}{}
 	\begin{algorithmic}%[1]
 		\renewcommand{\thealgorithm}{} 
 		\Function{$\tilde{\gamma } \,$}{}%{Fiber \& system \, parameters}
 		
 		\Comment{{\color{blue}  $\tilde{\gamma }$ calculation}}

   			\\\hrulefill
		    \State {\color{cyan} \bf Input variables}
   			\\\hrulefill
   			
		    \State \hspace{1 cm} {\color{red} \bf  Fiber parameters}
			\\\hrulefill
 		
 			\State $a_k $ 	\Comment{{\color{blue} Attenuation coefficient}} 
 		
 			\State $\beta _{2_k} $ \Comment{{\color{blue} GVD parameter}} 
 		
 			\State $\gamma_k$ \Comment{{\color{blue} nonlinear coefficient}}
 		
 			\State  $\ell _{s_k} $ \Comment{{\color{blue} Segment length}}
 			
 			\Statex

			\\\hrulefill
 		    \State \hspace{1 cm} {\color{red} \bf  System parameters}
			\\\hrulefill

		\State 	$\ell _{s} $ \Comment{{\color{blue} Span length}}
			
		\State	$L_{tot} $ \Comment{{\color{blue} Link length}}
			
		\State	$R_{s}$ \Comment{{\color{blue} Symbol rate}}

		\State	$N_{ch}$ \Comment{{\color{blue} Number of WDM channels}}

		\State	$N_{f}$ \Comment{{\color{blue} Number of fiber segments per span}}
			
		\State	$F_{A}$ \Comment{{\color{blue} EDFA noise figure}}

 			\\\vspace{0.1cm}
   			\\\hrulefill
			\State {\color{cyan} \bf Main code}
			\\\hrulefill
 		
 		\State \Comment{{\color{blue}Lower end of the integral }}
 		\State
 		
 		\State $sum \gets \kappa \left[\ln \left(\frac{\zeta _{0} }{\delta } \right)\int _{0}^{\delta }\xi  (\zeta )d\zeta +\delta \xi (0)\right] $
 		\State
 		\State $N_{int} \gets \left\lceil {(\zeta _{0}-\delta) }/{\pi } \right\rceil$ \Comment{{\color{blue} Number of $\pi$-intervals }}
 		\State
 		\While{$1\leq n\leq N_{int}$}
 		\State
 		\State $sum \gets sum + \tilde{\gamma }=\kappa \int _{(n-1)\pi}^{n \pi}\ln \left(\frac{\zeta _{0} }{\zeta } \right)\xi \left(\zeta \right)d\zeta $
 		\State
 		\EndWhile\label{euclidendwhile}
 		\State \textbf{return} $sum$ \Comment{{\color{blue} This is the nonlinear coefficient}}
 		\EndFunction
 	\end{algorithmic}
 \end{algorithm}

%\lstset{language=Mathematica}          % Set your language (you can change the language for each code-block optionally)
%
%\begin{lstlisting}[frame=single, mathescape=true]  % Start your code-block
%for i:=maxint to 0 do
%
%(* Assign values to system parameters *)
%
%$\ell _{s} \gets $ Span length (km)
%
%$L_{tot} \gets $ Link length (km)
%
%$R_{s} \gets $ Baud rate (Gbd)
%
%
%(* Assign values to fiber parameters *)
%
%for i:=1 to NumberOfSegments do
%begin
%{
%$a_k \gets$ Attenuation coefficient 
%
%$\beta _{2_k} \gets $ Group velocity dispersion (GVD) parameter 
%
%$\gamma_k \gets $ nonlinear coefficient
%
%$\ell _{s_k} \gets $ Segment length (km)
%}
%end;
%
%
%
%(* Define auxiliary functions *)
%
%$ \text{sinhc}(z) \df \frac{\sinh (z)}{z}$
%
%$\eta(\nu,\zeta) \df \Re\left[\frac{ \sin ^2(\zeta )+\sinh ^2(\nu )}{\zeta ^2+\nu ^2}\exp (-2 \nu )\right]$
%
%$\eta_c(z_1,z_2) \df \Re\left\{\text{sinhc}({z_ 2}) \text{sinhc}\left({z_1}^*\right) \exp \left[-({z_1}+{z_2})\right]\right\}$
%
%$\phi \left(\zeta \right)\df\frac{1}{N_{s}^{2} } \frac{\sin ^{2} (N_{s} \zeta )}{\sin ^{2} (\zeta )}$
%
%$f_{\phi }^{-1} (\beta _{2},\ell _{s} )\df 2\pi \sqrt{\left|\beta _{2} \right|\ell _{s} } $
%
%
%(* Comnpute nonlinear noise coefficient *)
%
%$\tilde{\gamma }=\kappa \int _{0}^{\zeta _{0} }\ln \left(\frac{\zeta _{0} }{\zeta } \right)\xi \left(\zeta \right)d\zeta$
%
%(* Output *)
%
%Write(Filename, results);
%
%\end{lstlisting}

 \subsection{Manakov equation}
 
\noindent   Before presenting our analytical calculations, in this subsection, we review some terminology and notation used throughout this paper. A full list of symbols is given at the end of the Appendix.

%Each fiber segment within a hybrid fiber span is characterized by its attenuation coefficient $\alpha _{k} ,$  GVD parameter $\beta _{2_{k} } ,$ and  nonlinear coefficient $\gamma _{k} .$ 

 We represent the optical signal by a two-dimensional complex vector $\vec{y}(z,t),$  whose components are the complex envelopes \cite{Proakis} of the signals along the $x,y$ states of polarization (SOPs). The vector components are functions of the position $z$ inside the fiber and the time $t.$ 

 We  adopt the shorthand notation of \cite{Mecozzi_JLT_12}, \cite{Ghazisaeidi:17}, for partial derivatives, where $\partial _{x} $ denotes partial differentiation with respect to the independent variable $x,$ $\partial _{x}^{2} $ denotes double partial differentiation with respect to $x,$ and so forth. Similarly, using Euler's notation, the symbol $D_{x} $ indicates a regular derivative with respect to $x.$ 
 
 In the remainder of this section, we discuss the formal derivation of a general expression for the nonlinear noise coefficient $\tilde{\gamma }$ in \eqref{eq:OSNRvsP}. %in order to clarify our notifications. 
  Our goal is to establish the connection among  disparate formalisms in previous publications \cite{Downie:17}, \cite{Miranda},  \cite{Al-Khateeb:18}, \cite{Krzczanowicz:19}. Portions of this formalism are taken from \cite{Poggiolini:12}, \cite{johannisson2013perturbation}, \cite{Poggiolini:14}, with changes in notation. 

 Based on Agrawal's derivation \cite{agrawalNL} but using the  engineering convention for the Fourier transform \cite{Proakis}\footnote{For a time-domain signal $x(t)$ with spectrum $X(f)$, the direct Fourier transform is defined as $X(f)=\int_{-\infty}^\infty x(t)e^{-i 2\pi f t}\,dt $ and the inverse Fourier transform is defined as $x(t)=\int_{-\infty}^\infty X(f) e^{i 2\pi f t}\,df $.}, the Manakov equation can be written  as follows \cite{forestieri2005solving}:
\begin{equation} 
\begin{split}
\partial _{z} \vec{y}(z,t)+\frac{a\left(z\right)}{2} \vec{y}(z,t)-\frac{i\beta _{2} \left(z\right)}{2} \partial _{t}^{2} \vec{y}(z,t)\\=-i\overline{\gamma} \left(z\right)\left\| \vec{y}(z,t)\right\| ^{2} \vec{y}(z,t) , 
\end{split}
\label{eq:Manakov_equation} 
\end{equation} 
where we neglected third-order chromatic dispersion and the optical amplifier noise. Notice that $\overline{\gamma} \left(z\right)=\frac{8}{9}\gamma \left(z\right)$ \cite{Wai_JLT_96}.

The difference between the above form of the Manakov equation and the one used in the conventional nonlinear Gaussian noise model \cite{Poggiolini:12}, \cite{Poggiolini:14}  is that we considered variable coefficients $a\left(z\right),\beta _{2} \left(z\right),\gamma \left(z\right).$ Further down, we assume that $a\left(z\right),\beta _{2} \left(z\right),\gamma \left(z\right),$ are piecewise constant functions of distance to express the fact that each optical fiber segment of a hybrid span has different characteristics.

\subsection{Solution for harmonic waves \label{sec:Harmonic_waves}}

\noindent   Based on the Manakov equation \eqref{eq:Manakov_equation}, we will study the non-linear propagation of each spectral component of the launched optical signal through the optical fiber. Initially, we will assume that the  signal generated by the optical transmitter is pseudorandom \cite{Poggiolini:12}, \cite{Poggiolini:14}, \cite{Carena:14}, i.e., periodic in time with period $T_0$. Due to the periodicity of the optical signal, its spectrum is composed of discrete spectral lines.  Later in the paper, in Sec. \ref{sec:nonlinear_noise_variance}, we will increase the signal period to infinity to deal with continuous signal spectra.

Since the launched optical signal is periodic, it can be expanded into exponential Fourier series
\begin{equation}  
\vec{y}(z,t)=\sum _{n \in \mathbb{Z}}\vec{u}_{n} (z)e^{i\omega _{n} t}  ,
\label{eq:Fourier_series} 
\end{equation} 
where $f_0=1/T_0$ is the fundamental frequency, $f_n=n f_0$ are the Fourier harmonics, $\omega _{n}=2\pi f_n $ are the corresponding angular frequencies, and  $\vec{u}_{n} (z)$ are complex Fourier coefficients which are vector functions of position. Later on, we omit the limits of the  infinite summation in the Fourier series in \eqref{eq:Fourier_series}  to avoid clutter.

 Substituting expression (\ref{eq:Fourier_series}) for $\vec{y}(z,t)$ into the Manakov equation (\ref{eq:Manakov_equation})  gives
\begin{equation} 
\begin{split} 
\sum _{n}\left[D_{z} \vec{u}_{n} (z)+\overline{a}_{n} \left(z\right)\vec{u}_{n} (z)\right]e^{i\omega _{n} t}  \\=-i\overline{\gamma} \left(z\right)\sum _{i,j,k}\left[\vec{u}_{k}^{\dag } (z).\vec{u}_{i} (z)\right]\vec{u}_{j} (z)e^{i\left(\omega _{i} +\omega _{j} -\omega _{k} \right)t}, 
\end{split}
\label{eq:ManakovEquationFourierCoeffs}
\end{equation} 
where a dagger $\dag $ denotes the adjoint matrix and we set
\begin{equation}  
\overline{a}_{n} \left(z\right)\df\frac{1}{2} \left[a\left(z\right)+i\beta _{2} \left(z\right)\omega _{n}^{2} \right]. 
\label{eq:complex_attenuation_coefficient}
\end{equation} 

The Fourier coefficients of two equal functions are equal \cite{Rudin}. By equating the angular frequencies in the two sides of (\ref{eq:ManakovEquationFourierCoeffs})
\begin{equation}  
\omega _{n} =\omega _{i} +\omega _{j} -\omega _{k} , 
\label{eq:angular_frequencies}
\end{equation} 
as well as the corresponding Fourier coefficients, we obtain the following system of coupled, first-order, ordinary differential equations (ODEs)
\begin{equation}
\begin{split}  
D_{z} \vec{u}_{n} (z)+\overline{a}_{n} \left(z\right)\vec{u}_{n} (z)\\=-i\overline{\gamma} \left(z\right)\sum _{(i,j,k) \in \Omega_n }\left[\vec{u}_{k}^{\dag } (z).\vec{u}_{i} (z)\right]\vec{u}_{j} (z) , 
\end{split}
\label{eq:ODE_system}
\end{equation} 
for $n\in \mathbb{Z}.$

In \eqref{eq:ODE_system},  $\Omega_n $ denotes the set of all index triplets for combinations of $\omega _{i} ,\omega _{j} ,\omega _{k} $ that create nonlinear interference at angular frequency $\omega _{n} $ through four-wave mixing due to Kerr effect,
\begin{equation}
\Omega_n\df\left\{(i,j,k) \in \mathbb{Z}^3 : \omega _{n} =\omega _{i} +\omega _{j} -\omega _{k} \right\} .
\label{eq:Combination_set}
\end{equation}

\subsection{Perturbation theory \label{sec:pert_theory}}

 %\noindent If we plot the variation of the $Q$-factor as a function of the launched power in the fiber $P$ per wavelength in both polarizations, we observe that, initially, the $Q$-factor  increases linearly with launched power $P$ , then it reaches an optimal value $Q_{0} $, and eventually decreases asymptotically as an inverse quadratic function of power.

% Contemporary long-haul coherent optical communications systems operate at the optimum point of the $Q-P$ characteristic (i.e., at the maximum $Q$-factor value) located at the beginning of the nonlinear operating regime. In this region, the non-linear term in the Manakov equation is small relative to the linear terms and can be considered as a small perturbation. Under these conditions, the Manakov equation can be solved analytically using perturbation theory \cite{nayfeh}, \cite{bender2013advanced}.

\noindent  To formally apply perturbation theory to the problem at hand, we artificially insert a small parameter $\varepsilon $ into the right hand side (RHS) of \eqref{eq:ODE_system}  \cite{bender2013advanced}
\begin{equation}  
\begin{split}
D_{z} \vec{u}_{n} (z)+\overline{a}_{n} \left(z\right)\vec{u}_{n} (z)\\=-i\varepsilon \overline{\gamma} \left(z\right)\sum _{(i,j,k)\in \Omega_n }\left[\vec{u}_{k}^{\dag } (z).\vec{u}_{i} (z)\right]\vec{u}_{j} (z) , 
\end{split}
\label{eq:Modified_Manakov_equation}
\end{equation} 
where we assumed that $\varepsilon$ is small enough that the impact of nonlinear effects on the solution is a small perturbation. At the end of the calculation, we will again set  $\varepsilon =1$ to obtain an approximate  analytical solution of \eqref{eq:ODE_system}. The accuracy of this solution will be determined by comparing the agreement among numerical and analytical results for the performance of various long-haul coherent optical communications systems \cite{Roudas:2020}.

 We assume that the  solution of \eqref{eq:Modified_Manakov_equation} can be expressed in terms of power series of the small parameter $\varepsilon$ \cite{nayfeh}, \cite{bender2013advanced}
\begin{equation} \label{eq:power_series_perturbation} 
\vec{u}_{n} (z)=\sum _{k=0}^{\infty }\vec{u}_{nk} (z)\varepsilon ^{k}  , 
\end{equation} 
where the term $\vec{u}_{nk} (z)\varepsilon ^{k} $ denotes the $k-$th  order correction to the unperturbed solution $\vec{u}_{n0} (z)$.

By substituting (\ref{eq:power_series_perturbation}) into the modified Manakov equation (\ref{eq:Modified_Manakov_equation}) and equating coefficients of like powers of $\varepsilon ,$ we obtain the following system of uncoupled, first-order, ODEs:
\begin{itemize}
\item Unperturbed ODE:
\begin{equation}  
D_{z} \vec{u}_{n0} (z)+\overline{a}_{n} \left(z\right)\vec{u}_{n0} (z)=\vec{0},
\label{eq:unperturbed_ODE} 
\end{equation} 

\item ODE for the $m-$th order perturbation ($m\geq 1$):
\begin{equation}  
\begin{split}
D_{z} \vec{u}_{nm} (z)+\overline{a}_{n} \left(z\right)\vec{u}_{nm} (z)=-i\overline{\gamma}\left(z\right)\\ \times\sum _{(i,j,k)\in \Omega_n }\sum _{(i',j',k')\in \Psi_m }\left[\vec{u}_{k k'}^{\dag } (z).\vec{u}_{i i'} (z)\right]\vec{u}_{j j'} (z) , 
\end{split}
\label{eq:perturbation_ODEs}
\end{equation} 
\end{itemize}
where $\Psi_m$ denotes the set of non-negative integers ${i'} ,{j'} ,{k'} $ 
%that satisfy the relationship
\begin{equation}
\Psi_m\df\left\{(i',j',k')\in {\mathbb{N}_0}^3 : i'+j'+k'+ 1 = m \right\} .
\label{eq:Combination_set_perturbation_ODEs}
\end{equation}

In the following, we will retain only the first-order perturbation term,
\begin{equation}\label{eq:first_order_power_series_perturbation}  
\vec{u}_{n} (z)\simeq\vec{u}_{n0} (z)+\varepsilon \vec{u}_{n1} (z), 
\end{equation} 
where the ODE for the first-order perturbation is given by \eqref{eq:perturbation_ODEs} by setting $m=1$
\begin{equation} 
\begin{split}
D_{z} \vec{u}_{n1} (z)+\overline{a}_{n} \left(z\right)\vec{u}_{n1} (z)\\=-i\overline{\gamma} \left(z\right)\sum _{(i,j,k)\in \Omega_n }\left[\vec{u}_{k0}^{\dag } (z).\vec{u}_{i0} (z)\right]\vec{u}_{j0} (z) . 
\end{split}
\label{eq:perturbed_ODE} 
\end{equation} 

The unperturbed ODE (\ref{eq:unperturbed_ODE}) can be solved by separation of variables
\begin{equation} \label{eq:unperturbed_ODE_solution} 
\vec{u}_{n0} (z)=\vec{c}_{n0}e^{-\int _{0}^{z}\overline{a}_{n} \left(z'\right)dz' } , 
\end{equation} 
where $\vec{c}_{n0}$ is the Fourier coefficient of the $n-$th spectral component at the fiber input.

 Assume that the solution of (\ref{eq:perturbed_ODE}) can be written in a form similar to (\ref{eq:unperturbed_ODE_solution})
\begin{equation} \label{eq:hypothetical_perturbed_ODE_solution} 
\vec{u}_{n1} (z)=\vec{c}_{n1} (z)e^{-\int _{0}^{z}\overline{a}_{n} \left(z'\right)dz' } , 
\end{equation} 
where the complex coefficient $\vec{c}_{n1} (z)$ is a function of distance $z$ to allow for nonlinear coupling introduced by the Kerr effect.

 By substituting \eqref{eq:unperturbed_ODE_solution}, (\ref{eq:hypothetical_perturbed_ODE_solution}) into (\ref{eq:perturbed_ODE}), we obtain the simplified ODE
\begin{equation} \label{eq:perturbed_ODE_complex_amplitude} 
\begin{split}
D_{z} \vec{c}_{n1} (z)=-i\overline{\gamma} \left(z\right)\sum _{(i,j,k)\in \Omega_n }\left[\vec{c}_{k0}^{\dag } .\vec{c}_{i0} \right]\vec{c}_{j0} \\ e^{-\int _{0}^{z}\overline{a}_{ijk} \left(z'\right)dz' } , 
\end{split}
\end{equation} 
where we defined the complex attenuation coefficient
\begin{equation} \label{eq:complex_attenuation_coefficient_2} 
\overline{a}_{ijk} \left(z\right)\df\overline{a}_{i} \left(z\right)+\overline{a}_{j} \left(z\right)+\overline{a}_{k}^{*} \left(z\right)-\overline{a}_{n} \left(z\right). 
\end{equation} 
In (\ref{eq:complex_attenuation_coefficient_2}), star denotes the complex conjugate.

 By integrating both sides of the above ODE over a single span length, we obtain
\begin{equation} \label{eq:c_n1_single_span} 
\boxed{\vec{c}_{n1} \left(\ell _{s} \right)=-\frac{8i}{9}\sum _{(i,j,k)\in \Omega_n }\left[\vec{c}_{k0}^{\dag } .\vec{c}_{i0} \right]\vec{c}_{j0}  X_{ijk} \left(\ell _{s} \right)}, 
\end{equation} 
where we defined the complex constant $X_{ijk} \left(\ell _{s} \right),$ which is related to the four-wave mixing efficiency
\begin{equation} \label{eq:complex_constant_FWM_efficiency} 
\boxed{X_{ijk} \left(\ell _{s} \right)\df\int _{0}^{\ell _{s} }{\gamma} \left(z\right)e^{-\int _{0}^{z}\overline{a}_{ijk} \left(z'\right)dz' } dz }. 
\end{equation} 
To derive (\ref{eq:c_n1_single_span}), we assumed, as initial condition, that the complex amplitude of the nonlinear noise at the fiber input is zero $\vec{c}_{n1} =\vec{0}$.

\subsection{Investigation of the FWM term \label{sec:FWM_term_cases}}

\noindent Let's focus our attention on the complex constant  $X_{ijk} \left(\ell _{s} \right).$ By substituting (\ref{eq:complex_attenuation_coefficient}), \eqref{eq:angular_frequencies}, into (\ref{eq:complex_attenuation_coefficient_2}), the complex attenuation coefficient $\overline{a}_{ijk}$ can be rewritten as
\begin{equation} \label{18)} 
\overline{a}_{ijk} \left(z\right)=a\left(z\right)+i\Delta \beta _{ijk} \left(z\right), 
\end{equation} 
where $\Delta \beta _{ijk} \left(z\right)$ is the phase mismatch
\begin{equation} \label{eq:phase mismatch_initial} 
\Delta \beta _{ijk} \left(z\right) \df -\beta _{2} \left(z\right)\left(\omega _{i}-\omega _{k}\right) \left(\omega _{j}-\omega _{k}\right) .
\end{equation} 

We can then calculate the nonlinear noise  generated at the center of the WDM spectrum. Setting $n =0$ in \eqref{eq:angular_frequencies}, we have
\begin{equation}
\omega _{k} =\omega _{i} +\omega _{j} .
\end{equation} 

We can then drop the subscript $k$ from the notation and simplify the relationships of the complex attenuation coefficient and the phase mismatch   
\begin{subequations}
	\begin{align}
	\overline{a}_{ij} \left(z\right) &= a\left(z\right)+i\Delta \beta _{ij} \left(z\right),\label{eq:aij}\\
	\Delta \beta _{ij}\left(z\right) &= -\beta _{2} \left(z\right)\omega _{i} \omega _{j} .\label{eq:Deltabetaij}
	\end{align}
\end{subequations}

\newpage
\subsection{Special cases of the FWM term (single span)}

\noindent \paragraph{Special case 1: Uniform fiber spans}

 For one fiber type per span, the fiber attributes are constant functions 
\begin{subequations}
	\begin{eqnarray}
	a\left(z\right)&=&a , \\ 
\beta _{2} \left(z\right)&=&\beta _{2}, 
\\ \gamma \left(z\right)&=&\gamma  .
\end{eqnarray}
\end{subequations}

Consequently, the complex attenuation coefficient $\overline{a}_{ij}\left(z\right)$ in \eqref{eq:aij} and the phase mismatch $\Delta \beta _{ij} \left(z\right)$ in \eqref{eq:Deltabetaij} are expressed in the simplified forms
\begin{subequations}
	\begin{eqnarray}
\alpha (i,j)&\df& \overline{a}_{ij}=a+i\Delta \beta(i,j) ,  \\ \Delta \beta (i,j) &\df& \Delta \beta _{ij}=-\beta _{2} \omega _{i} \omega _{j} .
\end{eqnarray} 
\end{subequations}
%where we omitted the dependence of $\alpha$ and $\Delta \beta$ on $i,j$ to avoid clutter.

It is straightforward to calculate the integral \eqref{eq:complex_constant_FWM_efficiency}  in closed form
\begin{equation} \label{eq:X_ij} 
X_{ij} \left(\ell _{s} \right)={\gamma}  \frac{1-e^{-\alpha(i,j) \ell _{s} } }{\alpha(i,j) } =\hat{\gamma }\hat{L}_{\rm eff} (i,j) ,
\end{equation} 
where we defined the effective nonlinear coefficient
\begin{equation} \label{21)} 
\hat{\gamma }\df\gamma ,
\end{equation} 
and the complex effective length
\begin{equation} \label{eq:Lhat_ij} 
\hat{L}_{\rm eff} (i,j) \df\frac{1-e^{-\alpha (i,j) \ell _{s} } }{\alpha  (i,j)}  .
\end{equation} 

Notice that \eqref{eq:Lhat_ij} deviates from the traditional definition of the effective length \cite{Agrawal}. The rationale behind our choice is that it sets a pattern for the resulting formula \eqref{eq:X_ij}  for $X_{ij} \left(\ell _{s} \right)$ that can be generalized to the  case of multiple fiber segments per span.
 
\noindent \paragraph{Special case 2: Two fiber types per span}

 Consider the situation where there are two fiber types per span with lengths $\ell _{s_{1} } $ and $\ell _{s_{2} } =\ell _{s} -\ell _{s_{1} } ,$ respectively. The fiber parameters of the two segments can be represented as step functions of $z$ 
\begin{subequations}
	\begin{eqnarray}\label{eq:piecewise_attenuation_coefficient} 
a\left(z\right)&=&\left\{\begin{array}{ccc} {a_{1} } & {} & {0\le z\le \ell _{s_{1} } } \\ {} & {} & {} \\ {a_{2} } & {} & {\ell _{s_{1} } < z\le \ell _{s} } \end{array} , \right.  
\\ &&\nonumber\\\label{eq:piecewise_GVD_parameter} 
\beta _{2} \left(z\right)&=&\left\{\begin{array}{ccc} {\beta _{21} } & {} & {0\le z\le \ell _{s_{1} } } \\ {} & {} & {} \\ {\beta _{22} } & {} & {\ell _{s_{1} } < z\le \ell _{s} } \end{array} , \right.  
\\ &&\nonumber\\\label{eq:piecewise_nonlinear_coefficient}
\gamma (z)&=&\left\{\begin{array}{ccc} {\gamma _{1} } & {} & {0\le z\le \ell _{s_{1} } } \\ {} & {} & {} \\ {\gamma _{2} } & {} & {\ell _{s_{1} } < z\le \ell _{s} } \end{array}.\right.  
\end{eqnarray} 
\end{subequations}

By substituting \eqref{eq:piecewise_attenuation_coefficient}-\eqref{eq:piecewise_nonlinear_coefficient} into \eqref{eq:complex_constant_FWM_efficiency} and by integrating analytically, we obtain
\begin{equation} \label{eq:complex_constant_FWM_efficiency_two_segments} 
X_{ij}\left(\ell _{s} \right)=\hat{\gamma }_{1} \hat{L}_{\rm eff_{1} } (i,j) +\hat{\gamma }_{2}  (i,j)\hat{L}_{\rm eff_{2} }  (i,j) ,
\end{equation} 
where, for notational convenience, we defined the effective nonlinear coefficients 
\begin{subequations}
	\begin{eqnarray}
	\label{eq:effective_nonlinear_coefficientsegment1} 
\hat{\gamma }_{1} & \df & \gamma _{1} ,
\\
\label{eq:effective_nonlinear_coefficientsegment2}  
\hat{\gamma }_{2} (i,j) & \df & \gamma _{2}  e^{-\alpha _{1}  (i,j) \ell _{s_{1} } }  ,
\end{eqnarray} 
\end{subequations} 
and the  complex effective lengths
\begin{equation} \label{eq:complex_effective_lengths} 
\hat{L}_{{\rm eff}_{k} } (i,j) \df \frac{1-e^{-\alpha _{k}  (i,j)\ell _{s_{k} } } }{\alpha _{k} (i,j)}  .
\end{equation} 
for $k=1,2$. 

The complex attenuation coefficient $\alpha _{k}  (i,j)$ for the $k-$th fiber segment is expressed in terms of the (real) attenuation coefficient $a_k$ and the phase mismatch $\Delta \beta_k (i,j)$ as
\begin{subequations}
\begin{eqnarray}
\alpha_k (i,j)&\df& a_k+i\Delta \beta_k (i,j) ,  \\ \Delta \beta_k (i,j) &\df& -\beta _{2k} \omega _{i} \omega _{j} .
\end{eqnarray} 
\end{subequations}

\paragraph{Special case 3: Multiple fiber types per span}

 \noindent Generalizing the above formulas to the case of $N_{f} $ fiber segments per span, the complex FWM efficiency is written as a sum
\begin{equation} \label{eq:complex_FWM_efficiency_Nf} 
X_{ij}\left(\ell _{s} \right)=\sum _{k=1}^{N_{f} }\hat{\gamma }_{k} (i,j) \hat{L}_{{\rm eff}_{k} }  (i,j) . 
\end{equation} 

In \eqref{eq:complex_FWM_efficiency_Nf}, the complex nonlinear fiber coefficients are defined as
\begin{equation} \label{eq:complex_nonlinear_fiber_coefficients_recursion} 
\hat{\gamma }_{k}  (i,j) \df\gamma _{k}  e^{-\sum _{m=1}^{k-1}\alpha _{m}  (i,j)\ell _{s_{m} }  } , 
\end{equation} 
and the  complex effective lengths are defined as
\begin{equation} \label{eq:normalized_complex_effective_lengths_Nf} 
\hat{L}_{{\rm eff}_{k} } (i,j) \df\frac{1-e^{-\alpha _{k}  (i,j)\ell _{s_{k} } } }{\alpha _{k} (i,j)  } . 
\end{equation} 

\begin{proof}

\noindent	Since the values of the fiber attributes  $a_{k} ,$  $\beta _{2_{k} } ,$ and  $\gamma _{k}$ are different for different fiber segments but constant within each segment, we  break up the integration interval in \eqref{eq:complex_constant_FWM_efficiency} into $N_f$ subintervals of width $\ell _{s_{k}}$. More specifically, we partition the $z$ axis with a sequence of $N_f+1$ points $z_0,\ldots, z_{N_f}$, where $z_0=0$ and $z_{N_f}=\ell _s$. The $k-$th fiber segment has endpoints $z_{k-1}$, $z_k$ and length $\ell _{s_{k}}=z_k-z_{k-1}$.
	
	\begin{eqnarray}
	X_{ij} \left(\ell _{s} \right) &=& \int _{0}^{\ell _{s} }\gamma  \left(z\right)e^{-\int _{0}^{z}\bar{a}_{ij}  \left(z'\right)dz'} dz \nonumber \\ &=&\sum _{k=1}^{N_{f} }\int _{z_{k-1} }^{z_{k} }\gamma  \left(z\right)e^{-\int _{0}^{z}\bar{a}_{ij}  \left(z'\right)dz'} dz  \nonumber\\ &=&\sum _{k=1}^{N_{f} }\gamma _{k} e^{-\int _{0}^{z_{k-1} }\bar{a}_{ij}  \left(z'\right)dz'} \nonumber\\&&\hspace*{1cm}\int _{z_{k-1} }^{z_{k} }e^{-\int _{z_{k-1} }^{z}\bar{a}_{ij}  \left(z'\right)dz'} dz   \nonumber\\ &=&
	\sum _{k=1}^{N_{f} }\gamma _{k} e^{-\sum _{m=1}^{k-1}\int _{z_{m-1} }^{z_{m} }\bar{a}_{ij}  \left(z'\right)dz' } \nonumber \\ &&\hspace*{1cm}\int _{z_{k-1} }^{z_{k} }e^{-\bar{a}_{ij} (z) \left(z-z_{k-1} \right)} dz  \nonumber \\&=&\sum _{k=1}^{N_{f} }\gamma _{k} e^{-\sum _{m=1}^{k-1}\alpha _{m} (i,j)\int _{z_{m-1} }^{z_{m} }dz'  } \nonumber \\&&\hspace*{1cm}\int _{0}^{\ell _{s_{k} } }e^{-\alpha _{k} (i,j)z} dz \nonumber \\ &=&\sum _{k=1}^{N_{f} }\gamma _{k} e^{-\sum _{m=1}^{k-1}\alpha _{m} (i,j)\ell _{s_{m} }  } \frac{1-e^{-\alpha _{k} (i,j) \ell _{s_{k}}} }{\alpha _{k} (i,j)} \nonumber  \\  &=&\sum _{k=1}^{N_{f} }\hat{\gamma }_{k} (i,j)\hat{L}_{{\rm eff}_{k} } (i,j) .
	\end{eqnarray}

\end{proof}

\subsection{Identical spans with multiple fiber types per span\label{sec:multiple_identical_spans}} 

\noindent	 For $N_{s} $ equal-length spans of length $\ell _{s}$ with $N_f$ fiber types per span and lumped optical amplifiers between successive spans to compensate for fiber attenuation, the  complex FWM efficiency for the full system length is given by
\begin{equation} \label{eq:coherent_addition_equal_length_spans} 
\begin{split}
X_{ij}\left(N_{s} \ell _{s} \right)=X_{ij}\left(\ell _{s} \right)\frac{\sin \left[N_{s} \Delta \beta (i,j)\ell _{s} /2\right]}{\sin \left[\Delta \beta (i,j)\ell _{s} /2\right]} \\e^{-i\left(N_{s} -1\right)\Delta \beta (i,j)\ell _{s} /2} , 
\end{split}
\end{equation} 
where $\Delta \beta $ is  the average  propagation constant mismatch 
\begin{equation}
\Delta \beta (i,j)\df\ell _{s}^{-1} \sum _{k=1}^{N_{f} }\Delta \beta _{k} (i,j)\ell _{s_{k} }  ,
\label{eq:average_phase_mismatch_DeltaBeta}
\end{equation} 
or, equivalently, $\beta _{2}$ is  the average  GVD parameter
\begin{equation}
\beta_2 \df\ell _{s}^{-1} \sum _{k=1}^{N_{f} }\beta _{2_k} \ell _{s_{k} }  .
\label{eq:average_phase_mismatch}
\end{equation} 

 \begin{proof}
 	
 \noindent Consider a link composed of $N_s$ identical spans of length $\ell _{s}$. The fiber attributes $a(z) ,$  $\beta _{2}(z) ,$ and  $\gamma (z)$ are periodic functions with fundamental period $\ell _{s}$. Therefore, we can write
 \begin{subequations}
 \begin{eqnarray} 
 a (z)& = & a\left(z-m\ell _{s} \right) , \label{eq:periodic_attenuation}\\
 \beta_{2} (z)&=& \beta _{2}\left(z-m\ell _{s} \right) , \label{eq:periodic_phase_mismatch} \\
 \gamma (z)&=& \gamma\left(z-m\ell _{s} \right), \label{eq:periodic_nl_coeff}
 \end{eqnarray} 
  \end{subequations}
 for $m \in \mathbb{Z}$ and $z$, $z-m\ell _{s}\in [0,N_s \ell _{s}]$.
 
 It follows that the complex attenuation coefficients $\overline{a}_{ij}\left(z\right)$ are also periodic functions with  period $\ell _{s}$. Thus, we can write
 \begin{equation}
  \bar{a}_{ij} (z) = \bar{a}_{ij}\left(z-m\ell _{s} \right) ,
  \label{eq:periodic_complex_attenuation}
 \end{equation}
  for $m \in \mathbb{Z}$ and $z$ as above.
  
To calculate $X_{ij} \left(N_{s} \ell _{s} \right)$, we start from \eqref{eq:complex_constant_FWM_efficiency},  break up the integration interval into subintervals of width $\ell _{s}$, and use the periodicity of \eqref{eq:periodic_attenuation}--\eqref{eq:periodic_nl_coeff} and \eqref{eq:periodic_complex_attenuation} to obtain
% 	\begin{eqnarray} 
% 	X_{ij} \left(N_{s} \ell _{s} \right)&=&\int _{0}^{N_{s} \ell _{s} }\gamma   \left(z\right)e^{-\int _{0}^{z}\bar{a}_{ij}  \left(z'\right)dz'} dz \nonumber\\&=&\sum _{m=0}^{N_{s}-1 }\left\{\int _{m\ell _{s} }^{(m+1)\ell _{s} }\right.\gamma  \left(z-m\ell _{s} \right)\nonumber\\&&\hspace*{1cm} \left. e^{-\int _{m\ell _{s} }^{z}\bar{a}_{ij}  \left(z'-m\ell _{s} \right) dz'} dz\right\}\nonumber\\&&\hspace*{1cm}  e^{-\int _{0}^{m\ell _{s} }\bar{a}_{ij}  \left(z'\right)dz'}  .
% 	\end{eqnarray} 
 	 \begin{multline*}
 	X_{ij} \left(N_{s} \ell _{s} \right)=\int _{0}^{N_{s} \ell _{s} }\gamma   \left(z\right)e^{-\int _{0}^{z}\bar{a}_{ij}  \left(z'\right)dz'} dz \nonumber\\=\sum _{m=0}^{N_{s}-1 }\Bigg\{ \int _{m\ell _{s} }^{(m+1)\ell _{s} }\gamma  \left(z-m\ell _{s} \right) e^{-\int _{m\ell _{s} }^{z}\bar{a}_{ij}  \left(z'-m\ell _{s} \right) dz'} dz \Bigg\} \\ e^{-\int _{0}^{m\ell _{s} }\bar{a}_{ij}  \left(z'\right)dz'}  . 
 	\end{multline*}
 	
 By changing the integration variables for the integrals in the curly brackets, \eqref{eq:periodic_complex_attenuation}, we obtain
% 	\begin{eqnarray} 
% 		X_{ij} \left(N_{s} \ell _{s} \right)&=&\left\{\int _{0}^{\ell _{s} }\gamma  \left(z\right)e^{-\int _{0}^{z}\bar{a}_{ij}  \left(z'\right)dz'} dz\right\}\nonumber\\&&\hspace*{1cm}\sum _{m=0}^{N_{s}-1 }e^{-\int _{0}^{m\ell _{s} }\bar{a}_{ij}^{(p)}  \left(z'\right)dz'}   \nonumber\\  &=&X_{ij} \left(\ell _{s} \right)\nonumber\\ &&\sum _{m=0}^{N_{s}-1 }e^{-\sum _{q=0}^{m-1}\int _{q\ell _{s} }^{(q+1)\ell _{s} }\bar{a}_{ij}  \left(z'-q\ell _{s} \right)dz' }  . \nonumber 
%\end{eqnarray} 
\begin{multline*}
X_{ij} \left(N_{s} \ell _{s} \right)=\\ \left\{\int _{0}^{\ell _{s} }\gamma  \left(z\right)e^{-\int _{0}^{z}\bar{a}_{ij}  \left(z'\right)dz'} dz\right\} \sum _{m=0}^{N_{s}-1 }e^{-\int _{0}^{m\ell _{s} }\bar{a}_{ij}  \left(z'\right)dz'}   \\=X_{ij} \left(\ell _{s} \right)\sum _{m=0}^{N_{s}-1 }e^{-\sum _{q=0}^{m-1}\int _{q\ell _{s} }^{(q+1)\ell _{s} }\bar{a}_{ij}  \left(z'-q\ell _{s} \right)dz' }  . \nonumber 
\end{multline*} 
 		 	
 	By performing a change of integration variables for the integral in the exponent, we obtain	 	
 		 	\begin{equation}
 		 		X_{ij} \left(N_{s} \ell _{s} \right)=X_{ij} \left(\ell _{s} \right)\sum _{m=0}^{N_{s}-1 }e^{-m\int _{0}^{\ell _{s} }\bar{a}_{ij}  \left(z'\right)dz'}.
 		 	\end{equation}
 		 	
 Due to periodic amplification at the end of each span, we discard the real part of $\int _{0}^{\ell _{s} } \bar{a}_{ij}  \left(z'\right)dz'$ and replace the latter integral with $i \int _{0}^{\ell _{s} }\Delta \beta _{ij}  \left(z'\right)dz' $.

 	Using the definition \eqref{eq:average_phase_mismatch_DeltaBeta} yields	 
 	\begin{equation}
 	X_{ij} \left(N_{s} \ell _{s} \right)=X_{ij} \left(\ell _{s} \right)\sum _{m=0}^{N_{s}-1 }e^{-i m \Delta \beta (i,j)\ell _{s} }  	.
 	\label{eq:xij_interim}
 	\end{equation}
 	
 	Finally, by summing the geometric series in \eqref{eq:xij_interim}, we get	
 	 	\begin{eqnarray} 	
 	 	X_{ij} \left(N_{s} \ell _{s} \right)&=& X_{ij} \left(\ell _{s} \right)\frac{1-e^{-iN_{s} \Delta \beta (i,j)\ell _{s} } }{1-e^{-i\Delta \beta (i,j)\ell _{s} } } \nonumber\\&=&X_{ij}\left(\ell _{s} \right)\frac{\sin \left[N_{s} \Delta \beta (i,j)\ell _{s} /2\right]}{\sin \left[\Delta \beta (i,j)\ell _{s} /2\right]} \nonumber\\&&\hspace*{1cm}e^{-i\left(N_{s} -1\right)\Delta \beta (i,j)\ell _{s} /2} . \label{eq:geometric-series}
 	\end{eqnarray} \end{proof}
 
\subsection{Final formula for the nonlinear noise variance\label{sec:nonlinear_noise_variance}}

 \noindent In this subsection, a passage is made from discrete to continuous signal spectra when the period of the transmitted signal $T_0\rightarrow\infty$. In the following integrals, we substitute the dummy variables $f_1,f_2$ for the  frequency components $f_i$, $f_j$, and abandon the indices $i,j$, used so far to keep track of the frequencies in the discrete setting.
 
  We consider an aperiodic WDM PDM signal that results from the superposition of $N_{ch}$ wavelength channels modulated at symbol rate $R_s$. From \cite[eq. (18)]{Carena_JLT_12}, the nonlinear noise psd $G_{\rm NLI} \left(f\right)$ can be written as 
\begin{equation}
\begin{split}
G_{\rm NLI} \left(f\right) \cong \frac{{16}}{{27}}N_s^2\int_{ - \infty }^\infty  {\int_{ - \infty }^\infty  {G({f_1})G({f_2})} } \\G({f_1} + {f_2} - f)\xi \left( {{f_1} - f,{f_2} - f} \right)df_1^{}df_2^{},
\end{split}
\label{eq:GNLI}
\end{equation} 
where $G(f)$ is the psd of the transmitted PDM WDM signal and the integrand equals
\begin{equation}
\xi \left(f_{1} ,f_{2} \right)\df\phi \left(f_{1} ,f_{2} \right)\eta \left(f_{1} ,f_{2} \right).
\label{eq:xi_integrand}
\end{equation} 

The first factor in \eqref{eq:xi_integrand} is the normalized phased-array term,  defined as
\begin{equation}
\phi \left(f_{1} ,f_{2} \right)\df\frac{1}{N_{s}^{2} } \frac{\sin ^{2} \left[ N_{s} \Delta \beta \left(f_{1} ,f_{2} \right) \ell _{s} /2)\right]}{\sin ^{2} \left[ \Delta \beta \left(f_{1} ,f_{2} \right) \ell _{s} /2 \right]} ,
\end{equation} 
where $\Delta \beta $ is  the average  phase mismatch in \eqref{eq:average_phase_mismatch_DeltaBeta}.
%\begin{equation}
%\Delta \beta =\ell _{s}^{-1} \sum _{k=1}^{N_{f} }\Delta \beta _{k} \ell _{s_{k} }  .
%\end{equation} 
 
The second factor in \eqref{eq:xi_integrand} is the four-wave mixing efficiency, defined as
\begin{equation}
\eta \left(f_{1} ,f_{2} \right)\df\left|\sum _{k=1}^{N_{f} }\hat{\gamma }_{k}^{} \left(f_{1} ,f_{2} \right)\hat{L}_{{\rm eff}_{k} }\left(f_{1} ,f_{2} \right)  \right|^{2} ,
\end{equation}  
which is the continuous counterpart of $|X_{ij}\left(\ell _{s} \right)|^2$ (see \eqref{eq:complex_FWM_efficiency_Nf}).

\subsection{Ideal Nyquist WDM spectra with zero roll-off factor \label{sec:Ideal Nyquist WDM spectra}}

 \noindent Here, we assume ideal Nyquist WDM spectra with zero roll-off factor. Furthermore, we assume that the WDM signal is a superposition of an odd number $N_{ch}$ wavelength channels  with spacing $\Delta \nu = R_{s}$. The optical bandwidth of the WDM signal is 
 \begin{equation}
 B_0=N_{ch} R_{s}.
 \label{eq:optical_bandwidth}
 \end{equation} 
 
 Approximating the hexagonal integration region \cite{Poggiolini:12}, \cite{Poggiolini:14}, resulting from \eqref{eq:GNLI} by a square, the nonlinear noise coefficient for the central WDM wavelength channel, measured in a resolution bandwidth $\Delta \nu _{{\rm res}}$, is given by the double integral
\begin{equation}
\tilde{\gamma }\simeq\frac{16}{27} \frac{N_{s}^{2} \Delta \nu _{{\rm res}} }{R_{s}^{3} } \int _{-B_{0} /2}^{B_{0} /2}\int _{-B_{0} /2}^{B_{0} /2}\xi \left(f_{1} ,f_{2} \right)df_{1} df_{2}   .
\label{eq:gamma_tilde_ideal_Nyquist}
\end{equation}

\subsection{Single integral for ideal Nyquist spectra \label{sec:single_integral}}

% The nonlinear noise coefficient is given by the double integral
%\begin{equation}
%\tilde{\gamma }=\frac{16}{27} \frac{N_{s}^{2} \Delta \nu _{{\rm res}} }{R_{s}^{3} }  \int _{-B_{0} /2}^{B_{0} /2}\int _{-B_{0} /2}^{B_{0} /2}\xi \left(f_{1} ,f_{2} \right)df_{1} df_{2}   ,
%\end{equation}  

\noindent We shall use a transformation of variables and iterated integration to convert \eqref{eq:gamma_tilde_ideal_Nyquist} into a single integral.

To begin, since $\xi \left(f_{1} ,f_{2} \right)$ is an even function of $f_{1} ,f_{2}, $ we can reduce the region of integration to the upper right quadrant of the coordinate plane
\begin{equation}
\tilde{\gamma }=\frac{64}{27} \frac{N_{s}^{2} \Delta \nu _{{\rm res}} }{R_{s}^{3} } \int _{0}^{B_{0} /2}\int _{0}^{B_{0} /2}\xi \left(f_{1} ,f_{2} \right)df_{1} df_{2}   .
\label{eq:gamma_tilde_double_integral_first_quadrant}
\end{equation}  

The integrand $\xi \left(f_{1} ,f_{2} \right)$ depends only on the product of the integration variables $f_{1} f_{2} $ so it is beneficial to define a new integration variable $\zeta $ that is directly proportional to $f_{1} f_{2} $ 
\begin{equation}
\zeta \df \frac{\Delta \beta \ell _{s} }{2} =\frac{f_{1} f_{2} }{2f_{\phi }^{2} } ,
\end{equation}  
where $f_{\phi }^{} $ is the average phased-array bandwidth \cite{Chen:OpEx10} defined as
\begin{equation}
f_{\phi }^{-1} \df 2\pi \sqrt{\left|\beta _{2} \right|\ell _{s} } .
\label{eq:average_phased-array_bandwidth}
\end{equation}  

Then, we change the integration variables from $f_{1} ,f_{2} $ to $f_{1} ,\zeta .$ By holding  $f_{1}$ fixed and differentiating with respect to $f_{2} $, we obtain $
d\zeta ={f_{1} df_{2} }/({2f_{\phi }^{2} }),$ or, equivalently,
$
df_{2} =({2f_{\phi }^{2} }/{f_{1} }) d\zeta .
$  
Since the upper limit of the integral in $f_2$ is $f_{2} ={B_{0} }/{2}$, the upper limit  of the integral in $\zeta $ becomes
$\zeta ={f_{1} B_{0} }/({4f_{\phi }^{2} }).$ 

With these substitutions, we obtain
\begin{equation}
\tilde{\gamma }=\frac{128}{27} \frac{N_{s}^{2} \Delta \nu _{{\rm res}} }{R_{s}^{3} }  f_{\phi }^{2} \int _{0}^\frac{B_{0}} {2}\frac{df_{1} }{f_{1} }\left[ \int _{0}^\frac{f_{1} B_{0}}{4f_{\phi }^{2}}\xi \left(\zeta \right)d\zeta  \right] .
\end{equation}  

Finally, we change the order of integration to transform the double integral into a single integral. By changing the integration order, the range of $f_{1} $  becomes $[(4f_{\phi }^{2} \zeta )/B_{0} ,B_{0} /2].$ Now $\zeta $ is the integration variable of the outer integral. Its limits correspond to the total range of $\zeta $ over the integration region $[0,B_{0}^{2} /\left(8f_{\phi }^{2} \right)].$ Hence,
\begin{equation}
\begin{split}
\tilde{\gamma }=\frac{128}{27} \frac{N_{s}^{2} \Delta \nu _{{\rm res}} }{R_{s}^{3} } f_{\phi }^{2} \int _{0}^\frac{B_{0}^{2} }{8f_{\phi }^{2}}\xi \left(\zeta \right)d\zeta \left[\int_{{4f_{\phi }^{2} \zeta }/{B_{0} }}^{B_{0} /2}\frac{df_{1} }{f_{1} } \right]  .
\end{split}
\end{equation}  

The inner integral is elementary and can be calculated in closed-form. Therefore, the double integral  can be transformed into a single-integral
\begin{equation}
\tilde{\gamma }=\frac{128}{27} \frac{\Delta \nu _{res} }{R_{s}^{3} } N_{s}^{2} f_{\phi }^{2} \int _{0}^{B_{0}^{2} /(8f_{\phi }^{2} )}\ln \left(\frac{B_{0}^{2} }{8\zeta f_{\phi }^{2} } \right)\xi \left(\zeta \right)d\zeta  .
\label{eq:single_integral_interim_form}
\end{equation}  

The latter integral must be computed using numerical quadrature. 

\subsection{Useful auxiliary quantities}

\noindent  In this subsection, we shall define some useful auxiliary quantities that will enable us to rewrite the integrand of \eqref{eq:single_integral_interim_form} in a  more appropriate form    for computation.

Notice that $\hat{\gamma }_{k}$, $\hat{L}_{{\rm eff}_{k} }$ given by \eqref{eq:complex_nonlinear_fiber_coefficients_recursion}, \eqref{eq:normalized_complex_effective_lengths_Nf}, respectively, depend on the products $\alpha _{k} \ell _{s_{k}}$. We can substitute these products by new complex coefficients $x_{k}\df\alpha _{k} \ell _{s_{k}}$.

We can rewrite $\hat{\gamma }_{k}^{} $ as
\begin{equation}
\hat{\gamma }_{k}^{} \left(\zeta \right) =\gamma _{k}^{}  e^{-\sum _{m=1}^{k-1}x_{m}\left(\zeta \right)  } ,
\label{eq:gammahat}
\end{equation}  
and $\hat{L}_{{\rm eff}_{k} } $ as
\begin{equation}
\hat{L}_{{\rm eff}_{k} } \left(\zeta \right) = \ell _{s_{k} }\frac{1-e^{-x_{k} \left(\zeta \right)} }{x_{k} \left(\zeta \right)} ,
\label{eq:Lhat_eff}
\end{equation}  
where, as mentioned above, we defined the normalized power complex attenuation coefficients $x_{k} \left(\zeta \right)$ as
\begin{equation}
x_{k} \left(\zeta \right)  \df {\alpha }_{k}\left(\zeta \right) \ell _{s_{k} } =2\left[\nu _{k} +i\zeta _{k} \left(\zeta \right) \right].
\label{eq:x_k}
\end{equation}  

In \eqref{eq:x_k}, $\nu _{k} $ stands for the normalized electric field attenuation coefficient 
\begin{equation}
\nu _{k} \df a_{k} \ell _{s_{k} } /2,
\label{eq:nu}
\end{equation}  
and $\zeta _{k} \left(\zeta \right)$ for the normalized electric field phase shift
\begin{equation}
\zeta _{k} \left(\zeta \right) \df \frac{\Delta \beta _{k} \ell _{s_{k} } }{2} =\frac{f_{1} f_{2} }{2f_{\phi _{k} }^{2} } =\frac{f_{\phi }^{2} }{f_{\phi _{k} }^{2} }\zeta.
\label{eq:zeta_k}
\end{equation}  

Similar to \eqref{eq:average_phased-array_bandwidth}, $f_{\phi _{k} }$ in \eqref{eq:zeta_k} denotes the phased-array bandwidth for each fiber segment
\begin{equation}
f_{\phi _{k} }^{-1} \df 2\pi \sqrt{\left|\beta _{2k} \right|\ell _{s_{k} } } .
\end{equation}

To further simplify the notation in \eqref{eq:zeta_k}, we can define the multiplicative coefficients
\begin{equation}
\lambda _{k} \df \frac{f_{\phi }^{2} }{f_{\phi _{k} }^{2} } ,
\label{eq:lambda}
\end{equation}  
so that the  arguments $\zeta _{k}\left(\zeta \right)  $ can be rewritten in compact form as a function of $\lambda _{k} $ and $\zeta $ 
\begin{equation}
\zeta _{k}\left(\zeta \right)  \df \lambda _{k} \zeta .
\end{equation}

\subsection{Final formalism\label{sec:Final_formalism}}

\noindent  After these definitions, the formalism for calculating the nonlinear noise coefficient $\tilde{\gamma }$ can be rewritten in compact form. 

From \eqref{eq:single_integral_interim_form}, the nonlinear noise coefficient, measured in a resolution bandwidth $\Delta \nu _{{\rm res}}=R_s$, is  expressed as a single definite integral 
\begin{equation}
\boxed{\tilde{\gamma }=\kappa \int _{0}^{\zeta _{0} }\ln \left(\frac{\zeta _{0} }{\zeta } \right)\xi \left(\zeta \right)d\zeta}  ,
\label{eq:nonlinear_noise_coefficient_single_integral}
\end{equation}   
where we defined
\begin{equation}
\kappa \df \frac{128}{27} \frac{f_{\phi }^{2} }{R_{s}^{2} } N_{s}^{2} ,
\end{equation}   

\begin{equation}
\zeta _{0} \df \frac{B_{0}^{2} }{8f_{\phi }^{2} } .
\label{eq:zeta_0}
\end{equation}   

The efficiency function $\xi \left(\zeta \right)$  is written as a product 
\begin{equation}
\xi \left(\zeta \right)=\phi \left(\zeta \right)\eta \left(\zeta \right)
\end{equation}   
of the normalized phased-array term
\begin{equation}
\phi \left(\zeta \right)=\frac{1}{N_{s}^{2} } \frac{\sin ^{2} (N_{s} \zeta )}{\sin ^{2} (\zeta )} 
\end{equation}   
and  the four-wave mixing efficiency
\begin{equation}
\boxed{\eta \left(\zeta \right)=\left|\sum _{k=1}^{N_{f} }\hat{\gamma }_{k}^{} \left(\zeta \right)\hat{L}_{{\rm eff}_{k} } \left(\zeta \right) \right|^{2}} .
\label{eq:FWM_efficiency-general_formula}
\end{equation}   

\subsection{Improper integral\label{sec:improper_integral}}

\noindent We want to numerically evaluate the integral \eqref{eq:nonlinear_noise_coefficient_single_integral}, which we rewrite below without the coefficient $\kappa$
\begin{equation}
I=\int _{0}^{\zeta _{0} }\ln  \left(\frac{\zeta _{0} }{\zeta } \right)\xi (\zeta )d\zeta .
\label{eq:integral}
\end{equation}

This is an improper integral of the second kind since the integrand has a singularity at zero 
$\mathop{\lim }\limits_{\zeta \to 0} \left({\zeta _{0} }/{\zeta } \right)=\infty .$
 
In order to evaluate $I$, we split the integration interval into two sub-intervals
\begin{equation}
I=\int _{0}^{\delta }\ln  \left(\frac{\zeta _{0} }{\zeta } \right)\xi (\zeta )d\zeta +\int _{\delta }^{\zeta _{0} }\ln  \left(\frac{\zeta _{0} }{\zeta } \right)\xi (\zeta )d\zeta ,
\label{eq:integral_decomposition}
\end{equation}
where $\delta$ is in the vicinity of $\zeta =0.$

\begin{figure}[!htb]\centering
	\includegraphics[width=0.4\textwidth]{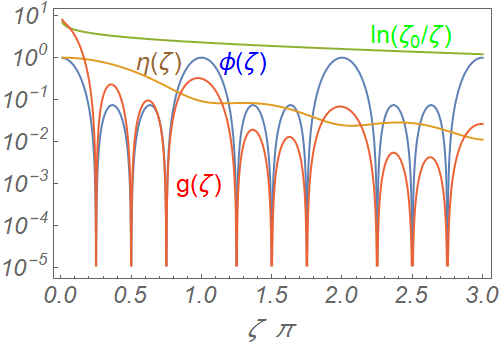}
	\caption{Sketches of $\ln  \left({\zeta _{0} }/{\zeta } \right)$ (in green), $\phi (\zeta )$  (in blue), $\eta (\zeta )/\eta (0 )$ (in brown), and their product  $g (\zeta )$ (in red). Conditions (for illustration purposes only): $N_s=4$, $\nu=1,$ $\zeta_0=10\pi,$ one fiber type per span.}
	\label{fig:IntegrandPlot}
\end{figure}

Some insight into the behavior of the integrand of \eqref{eq:integral} can be obtained from Fig. \ref{fig:IntegrandPlot}. 
As indicated by the red line, $g (\zeta )=\ln  \left({\zeta _{0} }/{\zeta } \right) \xi (\zeta )$ is oscillatory. The oscillation is mainly due to the phased-array factor $\phi (\zeta )$ (in blue), which is a periodic function with period $\pi .$ Principal maxima of unit height occur at integer multiples of $\pi .$ Between consecutive principal maxima (i.e., over a range of $\pi $) there are $N_{s} -1$ minima at multiples of  $\pi /N_{s} $ and $N_{s} -2$ subsidiary maxima approximately midway between successive minima. Thus, for a large number of spans $N_{s}$, the multiple-slit interference term $\phi (\zeta )$ rapidly varies over the integration region.  

For the first integral in \eqref{eq:integral_decomposition}, we can write
\begin{equation}
\begin{split}
\int _{0}^{\delta }\ln  \left(\frac{\zeta _{0} }{\zeta } \right)\xi (\zeta )d\zeta =\\
\ln  \left(\frac{\zeta _{0} }{\delta } \right)\int _{0}^{\delta }\xi (\zeta )d\zeta +
\int _{0}^{\delta }\ln  \left(\frac{\delta }{\zeta } \right)\xi (\zeta )d\zeta.
\end{split}
\label{eq:Taylor_expansion_integral_near_zero_initial}
\end{equation}

For the second integral in \eqref{eq:Taylor_expansion_integral_near_zero_initial}, taking the Taylor expansion of $\xi (\zeta )$ and integrating by parts, we obtain the following expression 
\begin{equation}
\begin{split}
\int _{0}^{\delta }\ln  \left(\frac{\delta }{\zeta } \right)\xi (\zeta )d\zeta=
\sum _{k=0}^{\infty }\frac{\delta ^{k+1} }{k!(k+1)^{2} }  {D_\zeta ^{k} \xi (0)} .
\end{split}
\label{eq:Taylor_expansion_integral_near_zero}
\end{equation}
An alternative expression is given by \eqref{eq:Jestf} in  Sec. \ref{sec:integral_singularity}.

Since $\eta \left(\zeta \right)$ (in brown in Fig. \ref{fig:IntegrandPlot}) is a slowly varying function of $\zeta ,$  in a small interval $\left[0,\delta \right]$, we can use the approximation $\xi \left(\zeta \right)\cong \eta \left(0\right)\phi \left(\zeta \right)$ in \eqref{eq:Taylor_expansion_integral_near_zero}. From L'H\^{o}pital's rule, $\phi (0) = 1$. The odd derivatives of $\phi \left(\zeta \right)$ are zero. The first few even derivatives $\partial _{\zeta }^{2k} \phi (0),$ for  $k\in \mathbb{Z},$ can be evaluated analytically
%\begin{eqnarray}
%	\phi (0) & = & 1 , \\&&\nonumber\\
%	\partial_{\zeta }^{2} \phi (0) & =& -\frac{2}{3} \left(N_{s}^{2} -1\right) \label{eq:second_derivative_phi} ,\\&&\nonumber\\
%	\partial _{\zeta }^{4} \phi (0) & =& \frac{8}{15} \left(2N_{s}^{4} -5N_{s}^{2} +3\right) .
%\end{eqnarray}
\begin{eqnarray}
\partial_{\zeta }^{2} \phi (0) & =& -\frac{2}{3} \left(N_{s}^{2} -1\right) \label{eq:second_derivative_phi}, \\&&\nonumber\\
\partial _{\zeta }^{4} \phi (0) &=&\frac{8}{15} \left(2N_{s}^{4} -5N_{s}^{2} +3\right).
\end{eqnarray}

%\begin{equation}
%\begin{array}{l@{}l}
%{\phi (0) &{} =1} \\ {\partial _{\zeta }^{2} \phi (0) &{} =-\frac{2}{3} \left(N_{s}^{2} -1\right)} \\ {\partial _{\zeta }^{4} \phi (0) &{} =\frac{8}{15} \left(2N_{s}^{4} -5N_{s}^{2} +3\right)}
%\end{array}  
%\end{equation}

For sufficiently small $\delta$, keeping only the zeroth-order term of the sum in the RHS of \eqref{eq:Taylor_expansion_integral_near_zero} is appropriate
\begin{equation}
\int _{0}^{\delta }\ln  \left(\frac{\zeta _{0} }{\zeta } \right)\xi (\zeta )d\zeta \cong \ln \left(\frac{\zeta _{0} }{\delta } \right)\int _{0}^{\delta }\xi  (\zeta )d\zeta +\delta \xi (0) .
\label{eq:series_impoper_integral}
\end{equation}

We can evaluate $\delta$ in \eqref{eq:series_impoper_integral} by imposing the condition that the zeroth-order term of the Taylor series in \eqref{eq:Taylor_expansion_integral_near_zero} should be much larger than the subsequent terms so that we can truncate the Taylor series to the zeroth-order term. Consequently, $\delta $ should satisfy the following inequality
\begin{equation}
\frac{\delta ^{3} }{2!3^{2} } \left|\partial _{\zeta }^{2} \xi (0)\right|\ll\delta \xi (0),
\end{equation}
which yields
\begin{equation}
\delta \ll\sqrt{\frac{2!3^{2} }{\left|\partial _{\zeta }^{2} \phi (0)\right|} } \stackrel{\eqref{eq:second_derivative_phi}}{=}\sqrt{\frac{3^{3} }{\left(N_{s}^{2} -1\right)} } \simeq \frac{3\sqrt{3} }{N_{s}^{} } .
\end{equation}
for $N_s \gg 1$.

Then, the two remaining integrals, 
\begin{equation}
\int _{\delta }^{\zeta _{0} }\ln  \left(\frac{\zeta _{0} }{\zeta } \right)\xi (\zeta )d\zeta ,
\end{equation}
and 
\begin{equation}
\int _{0}^{\delta }\xi  (\zeta )d\zeta ,
\end{equation}
in \eqref{eq:integral_decomposition} and \eqref{eq:Taylor_expansion_integral_near_zero_initial}, respectively, can be calculated numerically.

There are several numerical quadrature methods for highly oscillatory integrals \cite{olver2008numerical}. A rudimentary technique to uniformly sample the oscillatory integrand is Simpson's quadrature \cite{anton1995calculus}. The integration interval can be subdivided into subintervals of width $\pi /N_{s} $. We sample each subinterval $N_{n} $ times. Therefore, the distance between adjacent nodes is $\Delta =\pi /\left(N_{s} N_{n} \right).$

%For instance, the numerical evaluation is accurate for $\Delta =\pi /\left(10N_{s} \right),$$\delta =\pi /\left(5N_{s} \right).$

We have approximately $N_{\rm int}^{} =\left\lceil {\zeta _{0} }/{\pi } \right\rceil $ periods of $\phi \left(\zeta \right)$ in the interval $\left[0,\zeta _{0} \right].$ Then, we have $N_{\rm int}^{} N_{s} N_{n} $ nodes in the interval $\left[0,N_{\rm int}^{} \pi \right].$ Summing slices along the $\zeta$ axis can become cumbersome since \eqref{eq:optical_bandwidth} and \eqref{eq:zeta_0} give
\begin{equation}
N_{\rm int}^{} =\left\lceil \frac{N_{ch}^{2} R_{s}^{2} \left|\beta _{2} \right|\ell _{s} }{2} \right\rceil
, \end{equation}
so the number of periods of the multiple-slit interference term $\phi (\zeta )$  in the integration interval increases proportionally to the span length $\ell _{s} $ and quadratically with the number of WDM channels $N_{ch}^{} $ and the symbol rate $R_{s}^{} .$ 

Alternatively, the integral can be numerically evaluated using a commercial software tool like Mathematica \cite{mathematica}. For instance, the highly-oscillatory integral \eqref{eq:integral} can be accurately computed by partitioning the integration interval into $\pi$ subintervals, using the function NIntegrate in Mathematica with no options for each subinterval, and adding up the results.

\section{Additional bounds and approximations \label{sec:integral_truncation}}

\subsection{Truncation error}
% Looking back at (\ref{eq:nonlinear_noise_coefficient_single_integral}), to the effect that 
% $\tilde{\gamma }=\kappa \int _{0}^{\zeta _{0} }\ln \left(\frac{\zeta _{0} }{\zeta } \right)\xi \left(\zeta \right)d\zeta$

\noindent In this section, we estimate the tail contribution to the integral $I$ in (\ref{eq:integral}), as given by  
\begin{equation}
I(\mu,\zeta_0):=\int _{\mu}^{\zeta _{0} }\ln  \left(\frac{\zeta _{0} }{\zeta } \right)\xi(\zeta )d\zeta. 
\end{equation}
% For simplicity we take $\mu=(M+1)\pi$ and $\zeta_0=N \pi$ where $N>M$ are natural numbers.

To simplify formulas, we take the $\zeta$ cutoff value $\mu \in (0, \zeta_0)$ of the form $\mu=(M+1)\pi$, where $M$ is a  natural number. When considering large $\zeta_0$, little is lost
%\footnote{I could work around this need be.} 
in  assuming that $\zeta_0$ is also a multiple of $\pi$, $\zeta_0 = (N+1)\pi$. 
Our goal is to prove the following rigorous upper bound 
% \begin{align}
%   \label{final:est}
%   N_s I(\mu,\zeta_0) \leq&  \ \Gamma^2 \frac{1}{\sigma} \arccot(\frac{M \pi}{\sigma}) \ln(\frac{\zeta_0}{M\pi})
%   %\\                   \approx& \Gamma^2 \frac{1}{M \pi} \ln(\frac{\zeta_0}{M\pi}). % \qquad (\text{with $\mu=(M+1)\pi$})
% \end{align}
% where the right hand side
 \begin{subequations}
\begin{align}
\label{finalTail:est}
N_s I(\mu,\zeta_0) &\leq  \ \Gamma^2 \frac{1}{\sigma} \arccot(\frac{M \pi}{\sigma}) \ln(\frac{\zeta_0}{M\pi})
\\                   &\leq \ \Gamma^2 \frac{1}{M \pi} \ln(\frac{\zeta_0}{M\pi}), % \qquad (\text{with $\mu=(M+1)\pi$})
\end{align}
 \end{subequations}
where $\Gamma$, defined in (\ref{eq:enoncoeffDef}) ahead,  captures the non-linearity and $\sigma$ is the minimal adjusted attenuation, defined in (\ref{eq:adjustedAttDef}).  
The second inequality (based on $\arccot(x) \leq 1/x$ for $x>0$) reflects the $1/M$ asymptotics for large $M$. % (based on  $\arccot(x) \approx 1/x$). 
The upshot is that, when $\zeta_0$ is large and the oscillatory integrand in $I$ becomes problematic, $I$ can be typically approximated to within a few percent by its truncated version
\begin{equation}
I(0,\mu):=\int _{0}^{\mu}\ln  \left(\frac{\zeta _{0} }{\zeta } \right)\xi(\zeta )d\zeta ,
\end{equation} 
with a rigorous relative error $\epsilon_r$ bound  
\begin{equation}
\epsilon_r \leq
\frac{\Gamma^2}{M \pi N_s I(0,\mu)} \ln(\frac{\zeta_0}{M\pi}).
\end{equation}

\begin{proof}
	We derive (\ref{finalTail:est}) now.
	Substituting in (\ref{eq:FWM_efficiency-general_formula}) the definitions of the complex nonlinear coefficients $\hat{\gamma }_{k}^{}\left(\zeta \right)$ and complex effective lengths $\hat{L}_{{\rm eff}_{k} }\left(\zeta \right)$, given by \eqref{eq:gammahat} and \eqref{eq:Lhat_eff},
	%\footnote{JK PUT NEW LABELs} 
	yields a more detailed formula
	%\footnote{I like dividing  $\eta(\zeta)$ by $\ell_s^2$ and  using the $\frac{\beta_2}{\beta_{2_k}}$ over $\frac{\ell_{s_k}}{\lambda_k}$ in front of the $\gamma_k$, as in my original report.} 
	for the FWM efficiency  % \begin{equation}
	% \boxed{\eta \left(\zeta \right)=\left|\sum _{k=1}^{N_{f} }\hat{\gamma }_{k}^{} \left(\zeta \right)\hat{L}_{{\rm eff}_{k} } \left(\zeta \right) \right|^{2}}
	% \label{eq:FWM_efficiency-general_formula_finegrained}
	% \end{equation} 
	
	% \begin{equation}
	%   \eta_{\text{John}}(\zeta) : = \left|
	%     \sum_{k=1}^{n_f} \gamma_k \ell_{s_k} e^{-\sum_{m=1}^{k-1} 2 (\nu_m + i \lambda_m \zeta)} \frac{1-e^{-2 (\nu_k + i \lambda_k \zeta)}}{2 (\nu_k + i \lambda_k \zeta)}
	%   \right|^2  
	% \end{equation}
	
	\begin{align}\label{etaAdjusted:eq}
	\eta(\zeta) 
	=& \left|
	\sum_{k=1}^{N_f} \gamma_k \frac{\ell_{s_k}}{\lambda_k} e^{-\sum_{m=1}^{k-1} 2 \lambda_m (\sigma_m + i \zeta)} \frac{1-e^{-2 \lambda_k (\sigma_k + i \zeta)}}{2(\sigma_k + i \zeta)}
	\right|^2 \notag \\
	\end{align}
	where, to best capture the dependence on $\zeta$, we introduced normalized, chromatic dispersion-adjusted,  real attenuation coefficients for each fiber type: 
	\begin{equation}
	\label{eq:dAattDef}
	\sigma_k := \frac{\nu_k}{\lambda_k} = \frac{1}{2} \frac{\left|\beta_2\right|}{\left|\beta_{2_k}\right|} a_k \ell_s.
	\end{equation}
	
Recall that $\nu_k $ and
	$\lambda_k $ are given from \eqref{eq:nu} and \eqref{eq:lambda}, respectively.
	%$\lambda_k = \frac{f_\phi^2}{f_{\phi_k}^2} = \frac{\left|\beta_{2_k}\right|}{\left|\beta_2\right|}\frac{\ell_{s_k}}{\ell_s}$.)
	
	When the $\sigma_k$ do not vary dramatically between the fiber types, which is typically the case, a simple upper estimate of  $\eta(\zeta)$ can be made in terms of the minimum value 
	\begin{equation}
	\label{eq:adjustedAttDef}
	\sigma := \min\{ \sigma_k: \ k=1, \ldots, N_f\}.
	\end{equation}

	Specifically, combining the triangle inequality  with the monotonicity of $\frac{1+e^{-2\lambda_k \sigma_k}}{2 \sqrt{\sigma_k^2 + \zeta^2}}$
	as a function of $\sigma_k$ %$\frac{1}{\sigma_k^2 + \zeta^2}  \leq \frac{1}{\sigma^2 + \zeta^2}$
	gives 
	\begin{align}
	\label{eta:est}
	\eta(\zeta) 
	&\leq 
	\left( \sum_{k=1}^{N_f} \gamma_k \frac{\ell_{s_k}}{\lambda_k} e^{-\sum_{m=1}^{k-1} 2\lambda_m \sigma_m }\frac{1+e^{-2\lambda_k \sigma_k}}{2 \sqrt{\sigma_k^2 + \zeta^2}}    \right)^2  \notag
	\notag \\ &\leq  \left( \sum_{k=1}^{N_f}  \gamma_k \frac{\ell_{s_k}}{\lambda_k} e^{-\sum_{m=1}^{k-1} 2\lambda_m  \sigma }\frac{1+e^{-2\lambda_k \sigma}}{2}    \right)^2 \frac{1}{\sigma^2 + \zeta^2} \notag \\
	&= \Gamma^2 \frac{1}{\sigma^2 + \zeta^2} ,
	\end{align}
	where  we introduced 
	the worst-case (real) effective nonlinear  coefficient
	\begin{equation}
	\label{eq:enoncoeffDef}
	\Gamma :=  \sum_{k=1}^{N_f} \gamma_k \frac{\ell_{s_k}}{\lambda_k} e^{-\sum_{m=1}^{k-1} 2\lambda_m  \sigma }\frac{1+e^{-2\lambda_k \sigma}}{2}.
	\end{equation}

	One can think of $\Gamma$ as arising from a  hypothetical situation when the nonlinearities of individual fiber types only face the amount of attenuation of the lowest attenuation fiber and happen to all superpose constructively.
	%\footnote{POLISH the WORDING}  

	Switching attention to the phased-array term, we note that, multiplied  by $N_s$, it coincides with the Fej\'{e}r kernel  \cite{wiki:Fejer}
	\begin{equation}
	\label{eq:phaseArrayFejer}
	N_{s} \phi \left(\zeta \right)=\frac{1}{N_{s}} \frac{\sin ^{2} (N_{s} \zeta )}{\sin ^{2} (\zeta )}
	=  \sum_{j=-N_{s}+1}^{N_{s}-1} \left(1-\frac{|j|}{N_{s}}  \right)e^{2ij \zeta}, %  =  1 + 2 \sum_{j=1}^{n-1} \left(1-\frac{j}{n}\right) \cos(jx),
	\end{equation}
	which arises in the Ces\`aro summation of the Fourier series of $\pi$ periodic functions. In particular, it is $\pi$-periodic, non-negative, and  with  period average 
	\begin{equation}
	\frac{1}{\pi}\int_{0}^{\pi}  N_{s} \phi \left(\zeta \right)\, d\zeta =1.
	\end{equation}
	As a consequence, for any non-negative decreasing function $f(\zeta)$ and any $\zeta_1$, we have
	\begin{align}
	\label{period:est}
	\frac{1}{\pi} \int_{\zeta_1}^{\zeta_1+\pi}  N_s \phi(\zeta) f(\zeta) \, d\zeta  
	% \leq \int_{\zeta_1}^{\zeta_1+2\pi}  F_n(\zeta)  \max_{\zeta_0 \leq \zeta \leq \zeta_0+2\pi} f(\zeta)  \, d\zeta \leq 2\pi f(\zeta_0).
	%\leq \int_{\zeta_0}^{\zeta_0+\pi}  N_s \phi(\zeta)  f(\zeta_0)  \, d\zeta
	\leq f(\zeta_1).
	\end{align}

	% \begin{equation}
	%    I_n(\mu,\zeta_0) := N_s \int_\mu^{\zeta_0} \ln \left(\frac{\zeta _{0} }{\zeta } \right)\xi \left(\zeta \right)d\zeta
	% \end{equation}
	
	% \begin{equation}
	% \boxed{\tilde{\gamma }=\kappa \int _{0}^{\zeta _{0} }\ln \left(\frac{\zeta _{0} }{\zeta } \right)\xi \left(\zeta \right)d\zeta}  ,
	% \label{eq:nonlinear_noise_coefficient_single_integral}
	% \end{equation}
	
	%THE ESTIMATE
	%\bigskip
	
	%To estimate  $I(\mu,\zeta_0)$ let us 
	We are ready to estimate  $I(\mu,\zeta_0)$, or rather its $N_s$ rescaled version: 
	\begin{align}
	J(\mu,\zeta_0) := N_s I(\mu,\zeta_0) %= &  N_s \int_\mu^{\zeta_0} \ln \left(\frac{\zeta _{0} }{\zeta } \right)\xi \left(\zeta \right)d\zeta \notag \\
	= &  \int_\mu^{\zeta_0} N_s \phi(\zeta) \ln \left(\frac{\zeta _{0} }{\zeta } \right)\eta \left(\zeta \right)d\zeta.
	\end{align}

	% Still taking $\mu=(M+1)\pi$ pick a natural $N>M$ so that  $(N-1)\pi < \zeta_0 \leq N \pi$.
	Keeping in mind that  $\mu=(M+1)\pi$ and $\zeta_0=(N+1) \pi$ for natural $N>M$, combining  inequalities (\ref{eta:est}) and  (\ref{period:est}) and using the monotonicity of  $f(\zeta):= \ln(\frac{\zeta_0}{\zeta})\frac{\Gamma^2}{ \sigma^2 + \zeta^2 }$ gives 
	\begin{align}
	\label{eq:JestTail}
	J(\mu,\zeta_0) = & \int_\mu^{\zeta_0}  N_s \phi(\zeta) \ln(\frac{\zeta_0}{\zeta})\eta(\zeta)  \, d\zeta  \notag \\
	% =& \int_\mu^{\zeta_0}  F_n(2\zeta) \ln(\frac{\zeta_0}{\zeta}) \eta(\zeta) \, d\zeta  \notag \\
	\leq& \ \int_{(M+1) \pi}^{(N+1)\pi}    N_s \phi(\zeta)\ln(\frac{\zeta_0}{\zeta})  \frac{\Gamma^2}{ \sigma^2 + \zeta^2 } \, d\zeta  \notag \\
	\leq & \ \  \sum_{j=M+1}^{N}  \pi \ln(\frac{\zeta_0}{\pi j})   \frac{\Gamma^2}{ \sigma^2 + (\pi j)^2 }  \notag \\
	\leq & \ \int_{M\pi}^{N \pi} \ln(\frac{\zeta_0}{\zeta})  \frac{\Gamma^2}{ \sigma^2 + \zeta^2 } \, d\zeta.
	\end{align}
	%Integration by parts yields [TO DO MODIFY TO ALLOW $\zeta_0 \neq N \pi$] , setting $\mu':=M \pi = \mu - \pi$ for brevity,  
	
	Upon setting $\mu':=M \pi = \mu - \pi$ for brevity, the last integral can be integrated by parts: 
	\begin{align}
	% & \int_{\mu}^{\zeta_0} \ln(\frac{\zeta_0}{\zeta}) \frac{1}{\sigma^2 + \zeta^2}  \, d\zeta \notag \\
	%=& \int_{\mu}^{\zeta_0} \ln(\frac{\zeta_0}{\zeta}) \left(\frac{1}{\sigma} \arctan(\frac{\zeta}{\sigma})\right)' \, d\zeta \notag \\
	%=& \ln(\frac{\zeta_0}{\zeta}) \left(\frac{1}{\sigma} \arctan(\frac{\zeta}{\sigma})\right)\bigg|_{\mu}^{\zeta_0}   + \int_{\mu}^{\zeta_0} \frac{1}{\zeta} \left(\frac{1}{\sigma} \arctan(\frac{\zeta}{\sigma})\right) \, d\zeta \notag \\
	& \int_{\mu'}^{\zeta_0} \ln(\frac{\zeta_0}{\zeta}) \frac{1}{\sigma^2 + \zeta^2}  \, d\zeta = - \ln(\frac{\zeta_0}{{\mu'}}) \left[\frac{1}{\sigma} \arctan(\frac{{\mu'}}{\sigma})\right]  \notag \\
	& +\int_{\mu'}^{\zeta_0} \frac{1}{\zeta} \left[\frac{1}{\sigma} \arctan(\frac{\zeta}{\sigma})\right] \, d\zeta .
	\end{align}
	
	By using the crude estimate $\arctan(x) \leq \pi/2$, the above expression cannot exceed 
	\begin{align}
	\label{incoh:est}
	%& \int_{\mu'}^{\zeta_0} \ln(\frac{\zeta_0}{\zeta}) \frac{1}{\sigma^2 + \zeta^2}  \, d\zeta \notag \\
	%\ldots \leq&
	- \ln&\left(\frac{\zeta_0}{{\mu'}}\right) \left[\frac{1}{\sigma} \arctan(\frac{{\mu'}}{\sigma})\right]  + \frac{\pi}{2\sigma} \int_{\mu'}^{\zeta_0} \frac{1}{\zeta}  \, d\zeta \notag \\
	=&   \frac{1}{\sigma} \left[\pi/2 - \arctan(\frac{{\mu'}}{\sigma}) \right] \ln(\frac{\zeta_0}{{\mu'}}) \notag \\
	=&   \frac{1}{\sigma} \arccot(\frac{{\mu'}}{\sigma}) \ln(\frac{\zeta_0}{{\mu'}}).
	% \approx&  \frac{1}{{\mu'}} \ln(\frac{\zeta_0}{{\mu'}}).
	\end{align}
	
	Looking back at (\ref{eq:JestTail}), we have shown that $J(\mu,\zeta_0) \leq    \Gamma^2 \frac{1}{\sigma} \arccot(\frac{{\mu'}}{\sigma}) \ln(\frac{\zeta_0}{{\mu'}})$, which is the promised (\ref{finalTail:est}). 
\end{proof}

\subsection{Integral value in the vicinity of singularity\label{sec:integral_singularity}}

\noindent We will show that the part of the integral $I(0,\delta)$ contributed by a small interval $(0,\delta)$ in the neighborhood of zero is given by
\begin{align}
I(0,\delta)  = & \int _{0}^{\delta}\ln  \left(\frac{\zeta _{0} }{\zeta } \right)\xi(\zeta )d\zeta. \notag  \\
\simeq &\frac{\eta(0)}{N_s} \ln(\frac{\zeta_0}{\delta})\left[\delta+\sum_{j=1}^{N_s-1} \left(1-\frac{j}{N_s}  \right)\sin(2j \delta)\right]\notag \\&+\frac{\eta(0)}{N_s}\left[\delta+\sum_{j=1}^{N_s-1}\left(1-\frac{j}{N_s}  \right)\text{Si}(2j \delta)\right],
\label{eq:Jestf}
\end{align}
where $\text{Si}(x)$ denotes the {\it sine integral}: 
$$
\text{Si}(x) := \int_0^{x}  \frac{\sin(t)}{t}\, dt . %\leq \min\{\zeta, \pi/2 + \text{Gibbs overshoot} \}
%\leq \min\{\zeta, 2\}.
$$

The estimate \eqref{eq:Jestf} is derived by assuming that $\delta$ is sufficiently small so that $\eta(\zeta)\simeq\eta(0)$. % A safe choice for $\delta$ should be $\delta \simeq 1/N_s^2$. 

%\medskip

\begin{proof}
	First, we rewrite the sum (\ref{eq:phaseArrayFejer}) in real form 
	\begin{equation}
	N_s \phi(\zeta) %=  \sum_{j=-n+1}^{n-1} \left(1-\frac{|j|}{n}  \right)e^{2ij\zeta}
	=  1 + 2 \sum_{j=1}^{N_s-1} \left(1-\frac{j}{N_s}\right) \cos(2j\zeta).
	\end{equation}
	
	Then, it is straightforward to show that
	\begin{equation}
	\int_{0}^x  N_s \phi(\zeta) \, d\zeta =  x + \sum_{j=1}^{N_s-1} \left(\frac{1}{j}-\frac{1}{N_s}  \right) \sin(j2x).
	\label{eq:qux_integral_Fejer}
	\end{equation}

	Next, we compute the following auxiliary integral 
	\begin{equation}
	K_{N_s}(\delta) \df \int_0^{\delta} \ln(\delta/\zeta) N_s \phi(\zeta) \, d\zeta .
	\label{eq:K_integral_df}
	\end{equation}
	
	Notice that $K_{N_s}(\delta)$ can be rewritten as a double integral
\begin{equation}
	K_{N_s}(\delta) =  \int_0^{\delta} \left[ \int_\zeta^{\delta} \frac{1}{s}  \, ds \right] N_s \phi(\zeta) d\zeta . \notag \\
\label{eq:K_integral_double}
\end{equation}

	By switching the order of integration, we obtain
	%\footnote{or one could use  integration by parts}  
	\begin{align}
	K_{N_s}(\delta)
	&=  \int_0^{\delta} \frac{1}{s}  \left[ \int_{0}^s  N_s \phi(\zeta) \, d\zeta \right] ds \notag \\
	&=  \int_0^{\delta}  \frac{1}{s}  \left[  s + \sum_{j=1}^{N_s-1} \left(\frac{1}{j}-\frac{1}{N_s}  \right) \sin(j2s) \right] \, ds \notag \\
	&=    \delta + \sum_{j=1}^{N_s-1} \left(\frac{1}{j}-\frac{1}{N_s}  \right) \int_0^{\delta}  \frac{\sin(j2s)}{s}\, ds         \notag \\
	%  &=    \delta + \sum_{j=1}^{N_s-1} \left(\frac{1}{j}-\frac{1}{N_s}  \right) \int_0^{\delta}  \frac{\sin(2js)}{2js}\, d(2js)         \notag \\
	&=    \delta + \sum_{j=1}^{N_s-1} \left(\frac{1}{j}-\frac{1}{N_s}  \right) \int_0^{2 j \delta}  \frac{\sin(t)}{t}\, dt         \notag  \\
	&=    \delta + \sum_{j=1}^{N_s-1} \left(\frac{1}{j}-\frac{1}{N_s}  \right) \text{Si}(2j \delta).       \label{eq:KNs} 
	\end{align}
	
	We can write
	%\footnote{Using the right bound by $1$ under the minimum, yields $K_{N_s}(\delta) \leq \delta + 2 \ln N_s$.}:	
	\begin{align}
	\label{eq:JestDerived}
	I(0,\delta)  = & \int _{0}^{\delta}\ln  \left(\frac{\zeta _{0} }{\zeta } \right)\xi(\zeta )d\zeta. \notag  \\
	= & \frac{1}{N_s}\int_0^{\delta}  \ln(\frac{\zeta_0}{\zeta})N_s \phi(\zeta) \eta(\zeta)  \, d\zeta  \notag \\
	\simeq & \frac{\eta(0)}{N_s}\left[ \ln(\frac{\zeta_0}{\delta}) \int_0^{\delta}  N_s \phi(\zeta)   \, d\zeta  \right. \notag \\ + & \left. \int_0^{\delta} \ln(\frac{\delta}{\zeta})  N_s \phi(\zeta)  \, d\zeta \right]  . 
	\end{align}

Substituting \eqref{eq:qux_integral_Fejer} and \eqref{eq:KNs} into \eqref{eq:JestDerived}, we arrive at expression \eqref{eq:Jestf}. 

\end{proof}

As a corollary of the above calculation, an upper bound can be found for  $I(0,\delta)$. Using that $\sin(j2x)\leq j2x$ (for $x>0$) in \eqref{eq:qux_integral_Fejer}, we have  
	\begin{equation}
\int_{0}^{\delta}  N_s \phi(\zeta) \, d\zeta \leq  \delta + 2 \delta \sum_{j=1}^{N_s-1} \left(1-\frac{j}{N_s}  \right) = N_s \delta.
\label{eq:qux_integral_Fejer2}
\end{equation}

From \eqref{eq:KNs}, since $\text{Si}(\zeta)\leq \min\{\zeta, 2\}$, we have the following inequality  
\begin{align}
K_{N_s}(\delta)
&\leq  \delta + \sum_{j=1}^{N_s-1} \left(\frac{1}{j}-\frac{1}{N_s}  \right) \min\{2j\delta, 2\}. 
\end{align}
%For small $\delta$ (when $\delta < 1/N_s$),  one naturally % uses the left bound under the minimum, which gives   

In particular, choosing $2j\delta$ under the minimum, we get 
\begin{align}
\label{smallJbound:eq}
K_{N_s}(\delta)
&\leq  \delta + 2 \delta  \sum_{j=1}^{N_s-1} \left(1-\frac{j}{N_s}  \right)  = N_s \delta .
\end{align}

Then, substituting \eqref{eq:qux_integral_Fejer2} and \eqref{smallJbound:eq} into \eqref{eq:JestDerived} and using \eqref{eta:est}, we obtain the following bound 
\begin{align}
\label{eq:Jest}
I(0,\delta) &\leq \frac{\Gamma^2}{\sigma^2} \left[\ln(\frac{\zeta_0}{\delta}) + 1 \right] \delta.
\end{align}

\section{Results and discussion}

\noindent In this section, we focus our attention on the optimal design of a typical transatlantic coherent optical communications system with  hybrid fiber spans composed of an experimental QSMF \cite{QSMFpatent:17_1}, \cite{QSMFpatent:17_2} and a commercially-available, ultra-low-loss, large-effective-area SMF  without any splice losses. We evaluate, both analytically and numerically, the performance of various fiber configurations per span. We check the agreement between the analytical model of the previous section and Monte Carlo simulation,  and we show that the proposed GN model is sufficient for the determination of the optimum fiber splitting ratio.

\subsection{System parameters}
\noindent  We assume that the point-to-point link has total length equal to 6,000 km and is composed of  100 km spans. Furthermore, we assume an ideal Nyquist WDM signal composed of  9 wavelength channels, each carrying 32 GBd PDM 16-QAM. The attenuation coefficient of the QSMF is 0.16 dB/km and of the SMF is 0.158 dB/km. The effective mode area of the fundamental mode for the QSMF is 250 \textmu m$^2$ and of the SMF is 112 \textmu m$^2$. The GVD parameter $\beta_2$   is -26.6 ps$^{2}$/km for both fiber types.  The  EDFA noise figure is 5 dB.

Launching light in the fundamental mode of an ideal, straight, perfectly-cylindrical QSMF results, in theory, in pure single-mode propagation without coupling to higher-order modes. In practice, however, there always exists random coupling from the fundamental mode to higher-order modes and vice versa because of fiber irregularities. This  leads to the generation and propagation of a multitude of copies of the signal waveform across the fiber link. Due to modal dispersion, these signal copies propagate at various group velocities and interfere, either constructively or destructively, with the main signal propagating on the fundamental mode. This effect is referred to as multipath interference (MPI) \cite{Sui:15}, \cite{Mlejnek}. For modeling the impact of MPI-induced crosstalk, we assume that the QSMFs under consideration exhibit weak coupling between the fundamental mode group LP$_{01}$ and the higher-order mode group LP$_{11}$. For engineering purposes, we assume that MPI can be modeled as independent, zero-mean, complex Gaussian noises with a good degree of accuracy. Then, the MPI coefficient $\tilde{\beta }$ in \eqref{eq:OSNRvsP} can be calculated using power coupled-mode theory  \cite{Mlejnek}.

\subsection{Monte Carlo simulation results}
\begin{figure}[!htb]
	\centering
	\includegraphics[width=0.45\textwidth]{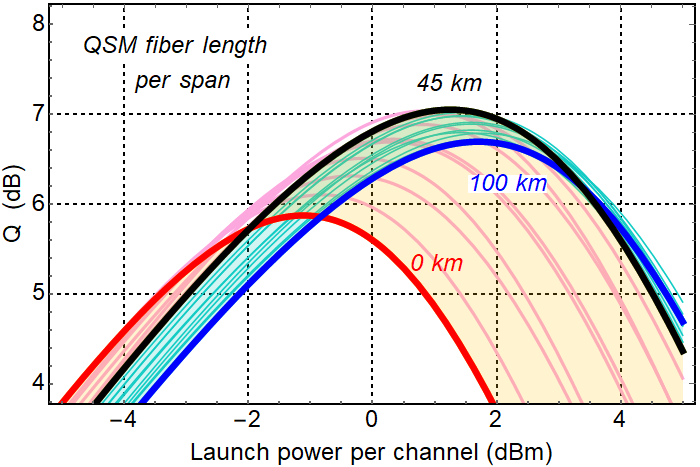}
	\caption{$Q$-factor as a function of the total launch power per channel for different QSMF lengths per span. (Conditions:  System length: 6,000 km, 100 km spans, QSMF effective mode area: 250 \textmu m${}^{2}$, SMF effective mode area: 112 \textmu m${}^{2}$. No MPI compensation. Lines: Fitting using \eqref{eq:OSNRvsP}.)} 
	\label{fig:QdBvsPplot_1800km_150kmspans_250_112um2_VariousCompensations_PhenomenologicalModel_AverageResults}
\end{figure}

\noindent Fig. \ref{fig:QdBvsPplot_1800km_150kmspans_250_112um2_VariousCompensations_PhenomenologicalModel_AverageResults} shows the variation of $Q-$factor as a function of the launch power per channel for  different fiber configurations, where the QSMF length per span is varied in the range 0--100 km in steps of 5 km. Lines represent least-squares fit of Monte Carlo simulation data with (1). To distinguish various simulation cases, we identify individual traces with different colors: fiber configurations with QSMF in the range 0--45 km are shown in pink and the remaining configurations for QSMF in the range 45--100 km per span are shown in cyan. We highlight the extreme cases for 0 km, 45 km, and 100 km using thick red, black, and blue lines respectively.

We observe that the optimum $Q-$factor increases as the QSMF length per span is increased up to 45 km. For QSMF segments longer than 45 km, the optimum $Q-$factor gradually declines, eventually reaching a 0.4 dB decrease at 100 km from the peak performance achieved at 45 km.

\subsection{Analytical model validation}
\begin{figure}[!ht]
	\centering
	
	\begin{subfigure}[b]{0.45\textwidth}
		\includegraphics[width=1\textwidth]{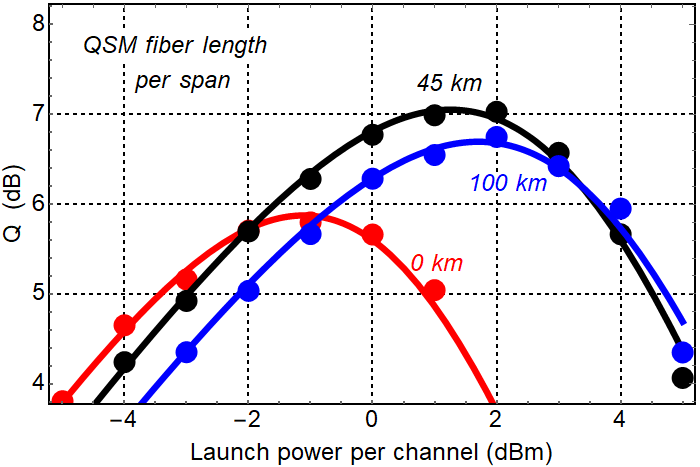}
		\caption{}
		\label{QdBvsP_6000km_100kmspans_250_112um2_0_MPI_Comp_Phenomenological_Fit}
	\end{subfigure}
	
	\vspace{1cm}
	
	%\vfill
	
	\begin{subfigure}[b]{0.45\textwidth}
		\includegraphics[width=1\textwidth]{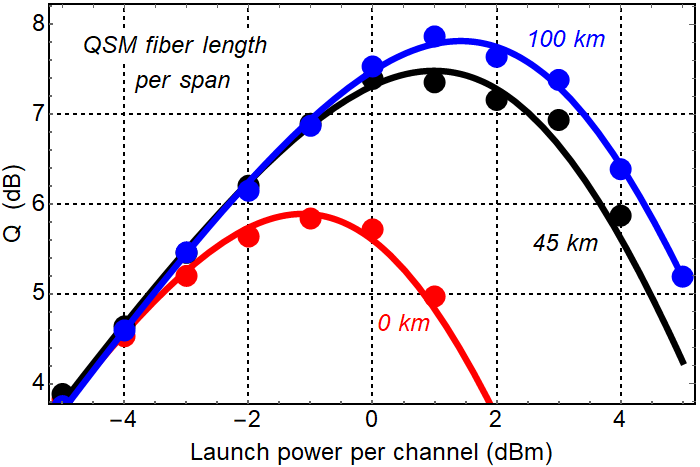}
		\caption{}
		\label{QdBvsP_6000km_100kmspans_250_112um2_100_MPI_Comp_Phenomenological_Fit}
	\end{subfigure}
	
	\caption{$Q$-factor as a function of the total launch power per channel.  (a) No MPI compensation; (b) 100\% MPI compensation. (Symbols: Points: Monte Carlo simulations; Lines: Fitting using \eqref{eq:OSNRvsP}.)}
	\label{fig:QdBvsP_6000km_100kmspans_250_112um2_Phenomenological_Fit}
\end{figure}

Since it is cumbersome to display both the analytical model and the numerical data on the same graph, we shall hereafter focus only on the extreme cases for 0 km, 45 km, and 100 km of the above graph.
Then, we replot the $Q-$factor against the average launch power as given by the Monte Carlo simulation for three different configurations of QSMF and SMF (Fig. \ref{fig:QdBvsP_6000km_100kmspans_250_112um2_Phenomenological_Fit}), i.e., using only SMF (in red), only QSMF (in blue), and a 45/55 mix of QSMF and SMF (in black).
We distinguish two cases:
\begin{enumerate}%[label=(\alph*)]
	\item	0\% MPI compensation at the coherent receiver (Fig. \ref{QdBvsP_6000km_100kmspans_250_112um2_0_MPI_Comp_Phenomenological_Fit}): The optimum $Q-$factor increases from 5.9 dB for SMF to 7 dB for 45/55 mix of QSMF and SMF and then drops to 6.7 dB for QSMF only. In this specific case, the optimum $Q-$factor is maximized with the use of 45 km QSM  fiber per span. The $Q-$factor improvement for using hybrid fiber spans compared to using SMF exclusively is 1.2 dB.
	\item	100\% MPI compensation (Fig. \ref{QdBvsP_6000km_100kmspans_250_112um2_100_MPI_Comp_Phenomenological_Fit}): The optimum $Q-$factor increases from 5.9 dB for SMF, to 7.5 dB for 45/55 mix of QSMF/SMF, to 7.8 dB for QSMF only. In this specific case, the optimum $Q-$factor is maximized with the exclusive use of QSMF per span. The $Q-$factor improvement for using only QSMF as opposed to using SMF exclusively is 1.9 dB.
\end{enumerate}

The lines in Fig. \ref{fig:QdBvsP_6000km_100kmspans_250_112um2_Phenomenological_Fit} are obtained by least squares fitting of the numerical results using \eqref{eq:OSNRvsP}. Notice that the numerical results and the fitted lines agree extremely well and this is a strong indication that \eqref{eq:OSNRvsP} is indeed an accurate model. However,  the analytical calculation of $\tilde{\gamma }$ in \eqref{eq:OSNRvsP} from first principles using the proposed nonlinear GN model is rather inaccurate.

\subsubsection{$Q$ vs. $P$  curves}
\noindent
Next, we check the accuracy of the proposed nonlinear GN model against  Monte Carlo simulation. We will show that the proposed nonlinear GN model describes qualitatively the general shape of the simulated $Q$ vs. $P$ curves but it does not provide pointwise accuracy. Nevertheless, as we are going to see subsequently, despite its quantitative errors, the proposed nonlinear GN model is sufficient for a quick determination of the optimum fiber splitting ratio.

\begin{figure}[!htb]
	\centering
	\includegraphics[width=0.45\textwidth]{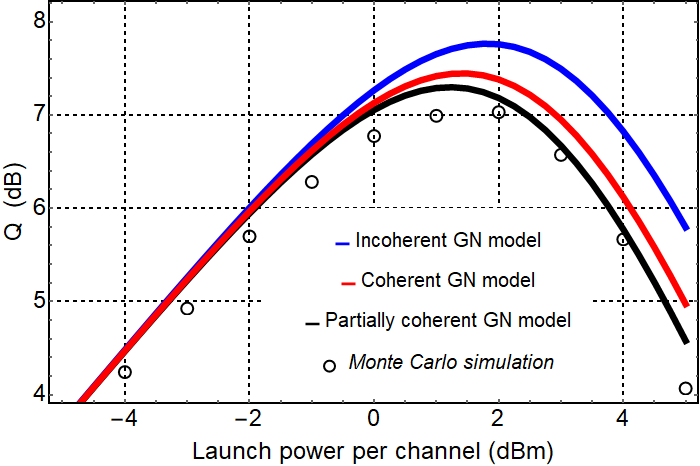}
	\caption{$Q$-factor as a function of the total launch power per channel for 45/55 QSMF/SMF mix. (Condition: No MPI compensation.)} 
	\label{fig:QdBvsPplot_fitting}
\end{figure}

As an illustration of the disagreement between the proposed nonlinear GN model and the simulation results, we replot from Fig.   \ref{QdBvsP_6000km_100kmspans_250_112um2_0_MPI_Comp_Phenomenological_Fit} the Monte Carlo simulation points (circles) describing the variation of the $Q-$factor as a function of the average launch power for the case of 45/55 QSMF/SMF mix in the absence of MPI compensation (Fig. \ref{fig:QdBvsPplot_fitting}).

On the same graph, we superimpose the incoherent nonlinear GN model with $\epsilon=0$ (in blue), the coherent nonlinear GN model (in red),  and  the partially-coherent nonlinear GN model with $\epsilon=0.15$ (in black). The analytical models based on coherent and incoherent addition deviate from the numerical results at relatively small launch powers. 
The peak deviation of the analytical curves in blue and red from the numerical results varies with the fiber attributes and system parameters. In this particular case, if we compare the values at the maxima, there is a mismatch of 0.4 dB between the coherent nonlinear GN model and the simulation. The discrepancy between the analytical and numerical results can be remedied to some extent by using the partially-coherent nonlinear GN model with $\epsilon$ as a fitting parameter (black line).

%\subsection{Validity of the analytical model: peak $Q-$factor $Q_0$ vs. QSMF length ${\ell }_{s_{\mathrm{1}}}$}
%\noindent 

\subsubsection{Optimum $Q-$factor vs. QSMF length}
\begin{figure}[!htb]
	\centering
	\includegraphics[width=0.45\textwidth]{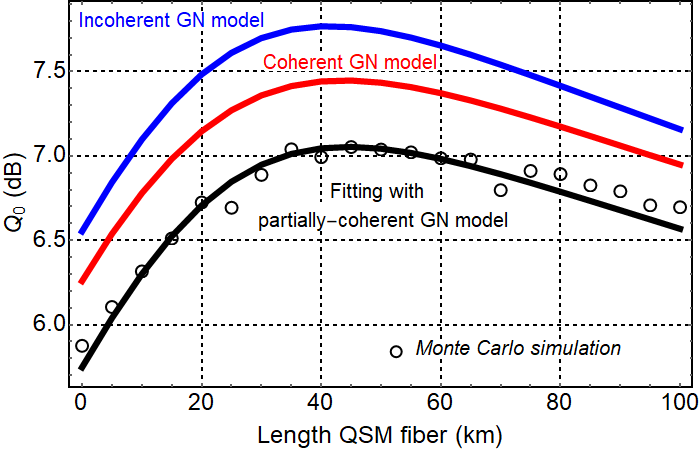}
	\caption{Peak $Q-$factor $Q_0$ vs. QSMF length ${\ell }_{s_{\mathrm{1}}}$ per span. (Condition: No MPI compensation.)} 
	\label{fig:Q0dBvsQSMFlengthplot_6000km_100kmspans_250_112um2_MC_inocherent_coherent_PoggioliniModel_v02}
\end{figure}

\noindent As another illustration of the validity of the analytical model, we examine the variation of the  peak $Q-$factor $Q_0$ as a function of the QSMF length  per span (Fig. \ref{fig:Q0dBvsQSMFlengthplot_6000km_100kmspans_250_112um2_MC_inocherent_coherent_PoggioliniModel_v02}). A major disagreement is apparent.
However, we notice that the optimum QSMF length, where the peak $Q-$factor $Q_0$ occurs, does not differ substantially from curve to curve.
The fact that we obtain essentially the same predictions for the optimum QSMF length from all the different variants of the analytical model is indicative of its usefulness.

%The numerical results together with theoretical predictions of the GN model are shown in Fig. \ref{fig:Q0dBvs.QSMFlengthplot_6000km_100kmspans_250_112um2_MC_inocherent_coherent_PoggioliniModel_v02}. 

\subsubsection{Optimum splitting ratio vs. MPI compensation}
\begin{figure}[!htb]\centering
	\includegraphics[width=0.45\textwidth]{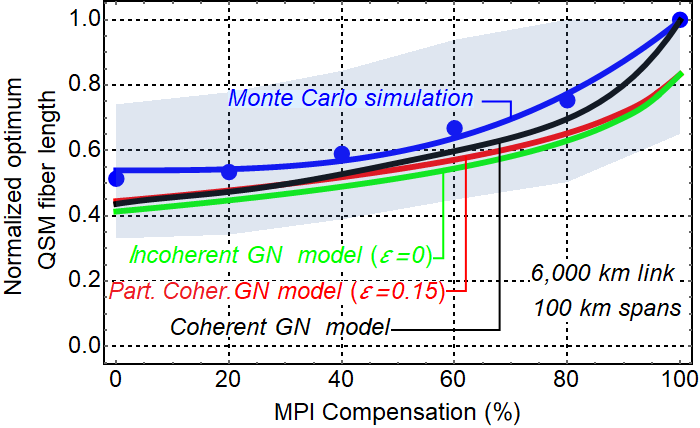}
	\caption{Variation of the optimal normalized QSMF length per span ${\ell }_{s_{\mathrm{1}}}/{\ell }_s$ as a function of the percentage of MPI compensation at the coherent optical receiver.} 
	\label{fig:SplittingRatioPlot_VariousCompensations_PhenomenologicalModel_01.png}
\end{figure}

\noindent
Fig. \ref{fig:SplittingRatioPlot_VariousCompensations_PhenomenologicalModel_01.png} shows a plot of the optimum splitting ratio vs. MPI compensation. The vertical axis  is normalized so that the span length ratio varies between zero and one.
Monte Carlo simulation data are represented by  blue points. The blue line shows a phenomenological model fit of the Monte Carlo simulation data. The blue shaded region around the blue line indicates  $\pm 0.1$ dB deviations from the optimum $Q-$values. The other  lines show the predictions of different variants of the modified nonlinear GN model.
As the MPI compensation increases, the ratio ${\ell }_{s_{\mathrm{1}}}/{\ell }_s$ increases to unity. The modified nonlinear GN model predictions are within the blue region.

Besides these validity checks, there are others presented by the authors at ECOC'17 \cite{Miranda} 
%and ONDM'19 \cite{Roudas_ONDM_19} 
for different fiber parameters that corroborate the current findings. Therefore, we believe that we have established the validity of the proposed analytical model for the practical determination of the optimum fiber splitting ratio per span. Henceforth, instead of numerically optimizing the lengths of the different fiber segments per span by solving the Manakov equation, which is a time consuming process, one can conveniently  resort to the analytical model.

\section{Summary}

\noindent Following the same methodology as the original nonlinear Gaussian noise model for uncompensated coherent optical communications systems with uniform fiber spans \cite{Poggiolini:12}, \cite{Poggiolini:14}, we provided here an up-to-date and, in some aspects,  improved derivation from first principles of an analytical relationship for the nonlinear Gaussian noise variance for hybrid fiber spans. 
Initially, we restated the full nonlinear Gaussian noise model in just 20 equations based on a synthesis of the literature. While the derivation presented here cannot claim to be fundamentally new, it is somewhat distinct from the one provided in the original publications on the nonlinear Gaussian noise model. 
Then, we derived new expressions for the nonlinear Gaussian noise variance for systems with multi-segment fiber spans. Even though these formulas were latent in \cite{Poggiolini:12}, \cite{Poggiolini:14}, and the most generic formalism \cite{semrau2018gaussian}, and variants of these formulas were published before, to the best of our knowledge, they were never proven before in their entirety. We hope to bring these formulas to broader attention. 
The most significant contribution of the current paper is the discussion of the accurate numerical evaluation of the definite integral for the nonlinear Gaussian noise variance, and the development of requisite estimates and asymptotics.
Finally, we performed extensive Monte Carlo simulation verification for a representative  transatlantic point-to-point link  of total length equal to 6,000 km with 100 km hybrid fiber spans, composed of an experimental QSMF and a commercially-available, ultra-low-loss, large-effective-area SMF without any splice losses. We showed that the modified nonlinear GN model is sufficiently accurate for the determination of the optimum fiber splitting ratio per span, yielding a system performance within $\pm 0.1$ dB  from the optimum $Q-$value.

\appendix

\subsection{Comparison of transformations of variables}

\noindent 		We investigate whether the use of  hyperbolic coordinates \cite{Poggiolini:12} can facilitate the evaluation of the double integral \eqref{eq:gamma_tilde_double_integral_first_quadrant} or not. We show that the integrand is simplified but the boundaries of the integration region become more complex. We conclude that hyperbolic coordinates offer no potential advantage compared to the transformation of variables proposed by the authors in Sec. \ref{sec:single_integral}.

\subsubsection{Hyperbolic coordinates}
\noindent This section is intended to show that the transformation of variables proposed by the authors is preferable to using hyperbolic coordinates as proposed by Poggiolini \cite{Poggiolini:14}, \cite{Poggiolini:17} because it is geometrically simpler and leads to the same final expression for the nonlinear noise coefficient in terms of a single definite integral in fewer steps.

\begin{figure}[!t]
	\begin{subfigure}{.25\textwidth}
	\centering
	\includegraphics[width=0.95\textwidth]{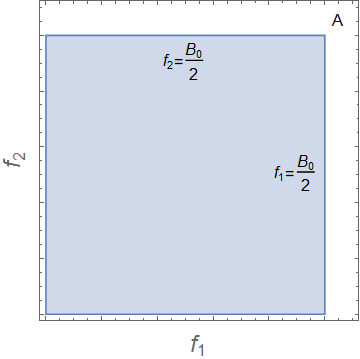}
	\caption{}
	\label{fig:squareIntegrationRegionPlot.png}
\end{subfigure}%
\begin{subfigure}{.25\textwidth}
	\centering
	\includegraphics[width=0.95\textwidth]{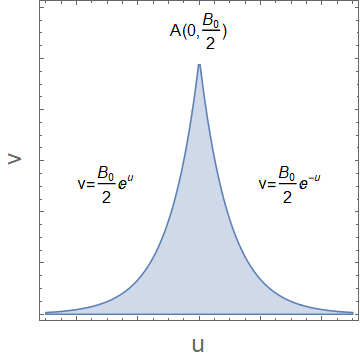}
	\caption{}
	\label{fig:circularTriangleHyperbolicCoordinatesIntegrationRegionPlot.png}
\end{subfigure}

\caption{Sketches of the integration regions (a) in the Cartesian $f_{1} f_{2} -$plane; (b) in the hyperbolic $uv-$plane.} 
\label{fig:IntegrationRegions}
\end{figure}

\noindent Consider the scaled version of the double integral \eqref{eq:gamma_tilde_double_integral_first_quadrant}
\begin{equation} \label{eq:initial_double_integral} 
I=\int _{0}^{B_{0} /2}\int _{0}^{B_{0} /2}\xi   \left(f_{1} f_{2} \right)df_{1} df_{2} . 
\end{equation} 
The region of integration  in the Cartesian $f_{1} f_{2} -$plane is a square of side $B_{0} /2,$ as shown in Fig. \ref{fig:squareIntegrationRegionPlot.png}.

Poggiolini \cite{Poggiolini:12} proposed to use hyperbolic coordinates \cite{wiki:Hyperbolic_coordinates} 
\begin{equation} \label{2)} 
u=\ln \sqrt{\frac{f_{1} }{f_{2} } } , 
\end{equation} 
\begin{equation} \label{3)} 
v=\sqrt{f_{1} f_{2} } . 
\end{equation} 

The inverse transform is \cite{wiki:Hyperbolic_coordinates} 
\begin{equation} \label{4)} 
\begin{array}{l} {f_{1} =ve^{u} ,\quad } \\ {f_{2} =ve^{-u} .} \end{array} 
\end{equation} 

Taking the Jacobian determinant yields
\begin{equation} \label{5)} 
df_{1} df_{2} =2vdvdu. 
\end{equation} 

The double integral can be rewritten
\begin{equation} \label{6)} 
I=2\iint\limits_{R}\xi \left(v^{2} \right)vdvdu,  
\end{equation} 
where $R$ is the region of integration in the hyperbolic $uv-$plane (shown in Fig. \ref{fig:circularTriangleHyperbolicCoordinatesIntegrationRegionPlot.png})
\begin{equation} \label{7)} 
R=\left\{\left(u,v\right)|u\in \mathbb{R},0\le v\le \min \left(\frac{B_{0} }{2} e^{u} ,\frac{B_{0} }{2} e^{-u} \right)\right\}. 
\end{equation} 

We notice that the integrand is an even function of $uv$ and that we can evaluate the double integral using only the first quadrant in the hyperbolic $uv-$plane. Carrying out first the integration in terms of $u,$ we have
\begin{equation} \label{8)} 
I=2\int _{0}^{B_{0} /2}\xi \left(v^{2} \right)dv^{2}  \left[ \int _{0}^{\ln \left[B_{0} /\left(2v\right)\right]}du \right], 
\end{equation} 
which yields
\begin{equation} \label{9)} 
I=2\int _{0}^{B_{0} /2}\ln \left(\frac{B_{0} }{2v} \right)\xi \left(v^{2} \right)dv^{2}  . 
\end{equation} 

With the additional change of variable
\begin{equation} \label{10)} 
\zeta =v^{2} , 
\end{equation} 
we finally obtain
\begin{equation} \label{eq:final_single_integral} 
I=\int _{0}^{B_{0}^{2} /4}\ln \left(\frac{B_{0}^{2} }{4\zeta } \right)\xi \left(\zeta \right)d\zeta  . 
\end{equation}

%\begin{tabular}{|p{2.2in}|p{2.2in}|} \hline 
%\includegraphics*[width=2.50in, height=2.49in, keepaspectratio=false]{image1} & \includegraphics*[width=2.50in, height=2.49in, keepaspectratio=false]{image2} \\ \hline 
%(a) & (b) \\ \hline 
%\end{tabular}

%\begin{figure}[!ht]
%	\centering
%	
%	\begin{subfigure}[b]{0.45\textwidth}
%		\includegraphics[width=1\linewidth]{Figs/squareIntegrationRegionPlot.png}
%		\caption{}
%		\label{fig:squareIntegrationRegionPlot.png}
%	\end{subfigure}
%	
%	\vspace{1cm}
%	
%	
%	\begin{subfigure}[b]{0.45\textwidth}
%		\includegraphics[width=1\linewidth]{Figs/circularTriangleHyperbolicCoordinatesIntegrationRegionPlot.png}
%		\caption{}
%		\label{fig:circularTriangleHyperbolicCoordinatesIntegrationRegionPlot.png}
%	\end{subfigure}
%	
%	\caption{Sketches of the integration regions (a) in the Cartesian $f_{1} f_{2} -$plane; (b) in the hyperbolic $uv-$plane.} 
%	\label{fig:QdBvsPplot_120kmspans}
%\end{figure}

\subsubsection{Proposed transformation}
\noindent Our proposed change of variables $\left(f_{1} ,f_{2} \right)\to \left(f_{1} ,\zeta \right)$ with $\zeta =f_{1} f_{2} $ gives the same result that was obtained in \eqref{eq:final_single_integral} in fewer steps and the integration region is simpler (i.e., it has a triangular shape---see Fig. \ref{fig:circularTriangleHyperbolicCoordinatesIntegrationRegionPlot.png}).

We rewrite \eqref{eq:initial_double_integral} 
\begin{equation} \label{12)} 
I=\int _{0}^{B_{0} /2}\int _{0}^{B_{0} /2}\xi   \left(f_{1} f_{2} \right)df_{1} df_{2} . 
\end{equation} 

Let
\begin{equation} \label{13)} 
\zeta =f_{1} f_{2} . 
\end{equation}

We evaluate the iterated integral
\begin{equation} \label{14)} 
I=\int _{0}^{B_{0}^{2} /4}\xi \left(\zeta \right)d\zeta  \left[\int _{2\zeta /B_{0} }^{B_{0} /2}\frac{df_{1} }{f_{1} } \right] , 
\end{equation} 
which simplifies to
\begin{equation} \label{15)} 
I=\int _{0}^{B_{0}^{2} /4}\ln \left(\frac{B_{0}^{2} }{4\zeta } \right)\xi \left(\zeta \right)d\zeta  . 
\end{equation}

%\begin{tabular}{|p{2.2in}|p{2.2in}|} \hline 
%\includegraphics*[width=2.50in, height=2.49in, keepaspectratio=false]{image3} & \includegraphics*[width=2.50in, height=2.49in, keepaspectratio=false]{image4} \\ \hline 
%(a) & (b) \\ \hline 
%\end{tabular}

%\begin{figure}[!ht]
%	\centering
%	
%	\begin{subfigure}[b]{0.45\textwidth}
%		\includegraphics[width=1\linewidth]{Figs/squareIntegrationRegionPlotSlice.png}
%		\caption{}
%		\label{fig:squareIntegrationRegionPlotSlice.png}
%	\end{subfigure}
%	
%	\vspace{1cm}
%	
%	
%	\begin{subfigure}[b]{0.45\textwidth}
%		\includegraphics[width=1\linewidth]{Figs/squareIntegrationRegionPlotSliceCoordinateTransfom.png}
%		\caption{}
%		\label{fig:squareIntegrationRegionPlotSliceCoordinateTransfom.png}
%	\end{subfigure}
%	
%	\caption{Sketches of the integration regions (a) in the Cartesian $f_{1} f_{2} $ plane; (b) in the Cartesian $f_{1} \zeta $ plane. The orange area among two consecutive contours $\zeta ,\zeta +\Delta \zeta $ on the left maps to the orange area on the right.} 
%	\label{fig:proposedTransform}
%\end{figure}
%
\begin{figure}[!t]
	\begin{subfigure}{.25\textwidth}
		\centering
		\includegraphics[width=0.95\textwidth]{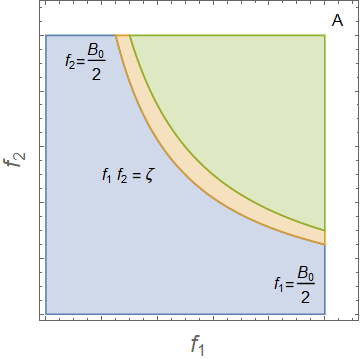}
		\caption{}
		\label{fig:squareIntegrationRegionPlotSlice.png}
	\end{subfigure}%
	\begin{subfigure}{.25\textwidth}
		\centering
		\includegraphics[width=0.95\textwidth]{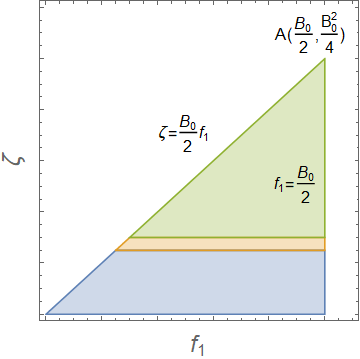}
		\caption{}
		\label{fig:squareIntegrationRegionPlotSliceCoordinateTransfom.png}
	\end{subfigure}
	
	\caption{Sketches of the integration regions (a) in the Cartesian $f_{1} f_{2}-$plane; (b) in the Cartesian $f_{1} \zeta-$plane. %The orange area among two consecutive contours $\zeta ,\zeta +\Delta \zeta $ on the left maps to the orange area on the right.
	} 
	\label{fig:proposedTransform}
\end{figure}

Integrating along $\zeta $ (Fig. \ref{fig:squareIntegrationRegionPlotSliceCoordinateTransfom.png}
) corresponds into slicing the integration region in the Cartesian $f_{1} f_{2} - $plane into hyperbolic segments given by $f_{1} f_{2} = $ constant (Fig. \ref{fig:squareIntegrationRegionPlotSlice.png}) and adding up the contribution of all slices.

Furthermore, Poggiolini \cite{Poggiolini:12}  suggested to restrict the domain of $\xi \left(f_{1} f_{2} \right)$ to $0\le f_{1} f_{2} \le c^{2} $  to accelerate numerical integration. We will see that this translates to truncating the tail of $\xi \left(\zeta \right)$ for $\zeta \ge c^{2} $ 
\begin{equation} \label{eq:truncated_integral_1} 
I\cong \int _{0}^{c^{2} }\ln \left(\frac{B_{0}^{2} }{4\zeta } \right)\xi \left(\zeta \right)d\zeta  . 
\end{equation} 

The appropriate upper limit $c^{2} $ can be calculated using the formalism in Sec. \ref{sec:integral_truncation}.

\subsection{Truncated integration region\label{sec:integral_truncation_2}}

\begin{figure}[!t]
	\begin{subfigure}{.25\textwidth}
		\centering
		\includegraphics[width=0.95\textwidth]{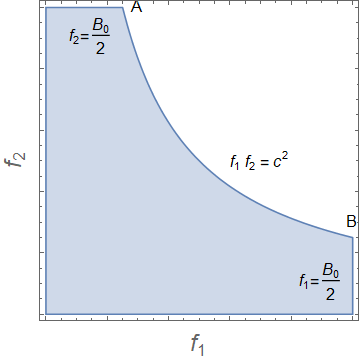}
		\caption{}
		\label{fig:truncatedAreaCartesianCoordinates.png}
	\end{subfigure}%
	\begin{subfigure}{.25\textwidth}
		\centering
		\includegraphics[width=0.95\textwidth]{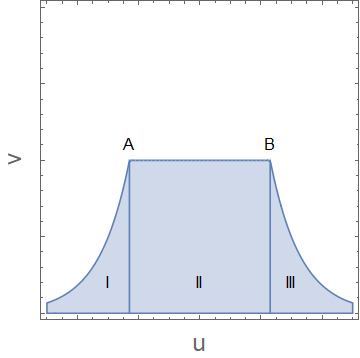}
		\caption{}
		\label{fig:truncatedAreaHyperbolicCoordinates.png}
	\end{subfigure}
	
	\caption{Sketches of the integration regions (a) in the Cartesian $f_{1} f_{2} -$plane; (b) in the hyperbolic $uv-$plane.} 
	\label{fig:truncatedRegion}
\end{figure}

\noindent We will evaluate \eqref{eq:initial_double_integral} over the truncated region shown in Fig. \ref{fig:truncatedAreaCartesianCoordinates.png}. We want to investigate whether it is beneficial to do so by using hyperbolic coordinates. We will show that the region of integration in the hyperbolic $uv-$plane (sketched in Fig. \ref{fig:truncatedAreaHyperbolicCoordinates.png}) is overly complicated compared to the region of integration obtained by the transformation of variables proposed by the authors (shown in blue in Fig. \ref{fig:squareIntegrationRegionPlotSliceCoordinateTransfom.png}).

%\begin{tabular}{|p{2.2in}|p{2.2in}|} \hline 
%\includegraphics*[width=2.50in, height=2.49in, keepaspectratio=false]{image5} & \includegraphics*[width=2.50in, height=2.49in, keepaspectratio=false]{image6} \\ \hline 
%(a) & (b) \\ \hline 
%\end{tabular}
%\begin{figure}[!ht]
%	\centering
%	
%	\begin{subfigure}[b]{0.45\textwidth}
%		\includegraphics[width=1\linewidth]{Figs/truncatedAreaCartesianCoordinates.png}
%		\caption{}
%		\label{fig:truncatedAreaCartesianCoordinates.png}
%	\end{subfigure}
%	
%	\vspace{1cm}
%	
%	
%	\begin{subfigure}[b]{0.45\textwidth}
%		\includegraphics[width=1\linewidth]{Figs/truncatedAreaHyperbolicCoordinates.png}
%		\caption{}
%		\label{fig:truncatedAreaHyperbolicCoordinates.png}
%	\end{subfigure}
%	
%	\caption{Truncated integration region (a) in the Cartesian $f_{1} f_{2} -$plane; (b) in the hyperbolic $uv-$plane.} 
%	\label{fig:truncatedRegion}
%\end{figure}

We assume that the region of integration in the $f_{1} f_{2} -$plane is enclosed by the lines
\begin{equation} \label{17)} 
f_{1} =\frac{B_{0} }{2} ,f_{2} =\frac{B_{0} }{2} ,f_{1} =0,f_{2} =0,f_{1} f_{2} =c^{2} . 
\end{equation} 
Then the region of integration in the $uv-$plane is limited by the lines
\begin{equation} \label{18)} 
v=\frac{B_{0} }{2} e^{u} ,v=\frac{B_{0} }{2} e^{-u} ,v=0,v=c. 
\end{equation} 
Points 
\[A=\left(\frac{2c^{2} }{B_{0} } ,\frac{B_{0} }{2} \right),B=\left(\frac{B_{0} }{2} ,\frac{2c^{2} }{B_{0} } \right), \] 
in the $f_{1} f_{2} -$plane correspond to points
\[A=\left(-\ln \left(\frac{B_{0} }{2c} \right),c\right),B=\left(\ln \left(\frac{B_{0} }{2c} \right),c\right), \] 
in the $uv-$plane. 

The region of integration in the $uv-$plane is divided into three subregions denoted by I-III in Fig. \ref{fig:truncatedAreaHyperbolicCoordinates.png}. The integral in subregion II is
\begin{equation} \label{19)}
\begin{split}  
I_{2} =2\int _{0}^{c}\xi \left(v^{2} \right)dv^{2}  \left[\int _{0}^{\ln \left[B_{0} /\left(2c\right)\right]}du\right] \\=2\ln \left(\frac{B_{0} }{2c} \right)\int _{0}^{c}\xi \left(v^{2} \right)dv^{2}  . 
\end{split}
\end{equation} 

With the additional change of variable
\begin{equation} \label{20)} 
\zeta =v^{2} , 
\end{equation} 
we finally obtain
\begin{equation} \label{21)} 
I_{2} =2\ln \left(\frac{B_{0} }{2c} \right)\int _{0}^{c^{2} }\xi \left(\zeta \right)d\zeta  . 
\end{equation} 

The two integrals for subregions I and III are identical. Their sum is 
\begin{equation} \label{22)} 
\begin{split} 
I_{1+3} =2I_{3} =2\int _{0}^{c}\xi \left(v^{2} \right)dv^{2}  \left[\int _{\ln \left[B_{0} /\left(2c\right)\right]}^{\ln \left[B_{0} /\left(2v\right)\right]}du\right] \\=2\int _{0}^{c}\ln \left(\frac{c}{v} \right)\xi \left(v^{2} \right)dv^{2} =\int _{0}^{c}\ln \left(\frac{c^{2} }{v^{2} } \right)\xi \left(v^{2} \right)dv^{2}   . 
\end{split} 
\end{equation} 

Let again $\zeta =v^{2} ,$ so that
\begin{equation} \label{24)} 
I_{1+3} =\int _{0}^{c^{2} }\ln \left(\frac{c^{2} }{\zeta } \right)\xi \left(\zeta \right)d\zeta  . 
\end{equation}

The final integral is
\begin{equation} \label{25)} 
\begin{split} 
I=I_{2} +I_{1+3} =\ln \left(\frac{B_{0}^{2} }{4c^{2} } \right)\int _{0}^{c^{2} }\xi \left(\zeta \right)d\zeta \\+\int _{0}^{c^{2} }\ln \left(\frac{c^{2} }{\zeta } \right)\xi \left(\zeta \right)d\zeta  .
\end{split}  
\end{equation} 
or, equivalently,
\begin{equation} \label{26)} 
I=\int _{0}^{c^{2} }\ln \left(\frac{B_{0}^{2} }{4\zeta } \right)\xi \left(\zeta \right)d\zeta  . 
\end{equation} 

The integral is identical to the one that we had tried to compute initially (see \eqref{eq:truncated_integral_1}). It turns out that the upper bound $c^{2}$ determines the contour at which we have to truncate the integrand. It is equal to the upper limit $\mu$ in  Sec. \ref{sec:integral_truncation}A. In other words, the physical meaning of $\mu$ in Sec. \ref{sec:integral_truncation}A is
$\mu =c^{2} =f_{1} f_{2}.$

%\section*{List of symbols}
%
%	\mbox{}
%	
%	\nomenclature{$c$}{Speed of light in a vacuum inertial frame}
%	\nomenclature{$h$}{Planck constant}
%	\nomenclature{$\beta$}{The second letter of the greek alphabet}
%	\nomenclature{$\alpha$}{The first letter of the greek alphabet}
%	
%	
%	\printnomenclature
	
	\section*{List of symbols}
	\addcontentsline{toc}{section}{Nomenclature}
	\begin{IEEEdescription}[\IEEEusemathlabelsep\IEEEsetlabelwidth{$V_1,V_2,V_3$}]
\item[$a$]  Attenuation coefficient.
\item[$A_{\text{eff}}$]  Mode effective area.
\item[$\beta_2$]  Group velocity dispersion (GVD) parameter.
\item[$D$]  Chromatic dispersion parameter.
\item[$\Delta\nu$]  Frequency spacing of WDM channels.
\item[$\ell _{s}$] Span length.
\item[$F_{A} $]  Amplifier noise figure.
\item[$G$]  Amplifier gain.
\item[$L$] Link length.
\item[$\lambda$]  Carrier wavelength of central WDM channel.
\item[$n_{2}$]  Nonlinear index coefficient.
\item[$N_{ch}$]  Number of wavelength channels.
\item[$N_{s}$] Number of spans.
\item[$N_f$] Number of fiber segments per span.
\item[$\vec{y}(z,t)$]  Complex envelope WDM PDM signal.
\item[$\partial _{x} $ ]  Partial derivative $\partial/\partial{x} $.
\item[$D_{x}$ ]  Regular derivative $d/d{x} $.
\item[$\overline{\gamma}$ ] Averaged nonlinear coefficient $\overline{\gamma}=\frac{8}{9}\gamma$.
\item[${\gamma}$ ]  Nonlinear coefficients.
\item[$\overline{a}_{n} $ ]  Complex attenuation coefficient.
\item[$\omega_{n} $ ]  Angular frequencies $\omega_{n} = 2 \pi f_n$.
\item[$ \vec{u}_{n} (z)$ ] Fourier coefficients.
\item[$ \Omega_n $ ] Set of index triplets for FWM combinations.
\item[$\varepsilon $ ] Perturbation parameter.
\item[$T_0$] Pseudorandom signal period.
\item[$f_0$] Pseudorandom signal fundamental frequency.
\item[$\vec{u}_{nk} (z)\varepsilon ^{k} $] $k-$th  order correction to the unperturbed solution $\vec{u}_{n0} (z)$
\item[$\Psi_m$] Set of index triplets for the ODE for the $m-$th order perturbation.
\item[ $\vec{c}_{n0}$] Complex envelope of the unperturbed Fourier coefficient of the $n-$th spectral component at the fiber input.
\item[ $\overline{a}_{ijk} \left(z\right)$] Complex attenuation coefficient.
\item[ $X_{ijk},X_{ij}$] Complex FWM efficiency.
\item[ $\vec{c}_{n1}$] Complex amplitude of the nonlinear noise.
\item[ $\Delta \beta _{ijk} \left(z\right)$] Phase mismatch.
\item[ $\Delta \beta$] Average  propagation constant mismatch.
\item[ $\hat{\gamma }$] Effective nonlinear coefficient.
\item[ $\hat{L}_{\rm eff}$] Normalized (i.e., dimensionless) complex effective length.
\item[ $\phi \left(\zeta \right)$] Normalized phased-array term.
\item[ $\eta \left(\zeta \right)$] Four-wave mixing efficiency.
\item[ $\tilde{a}$]  Amplified spontaneous emission (ASE) noise variance.
\item[ $\tilde{\beta }P$]  Multipath crosstalk variance.
\item[ $\tilde{\gamma }P^{3}$]  Nonlinear noise variance.
\item[ $\phi \left(f_{1} ,f_{2} \right)$]  Normalized phased-array term.
\item[ $\eta \left(f_{1} ,f_{2} \right)$]  Four-wave mixing efficiency.
\item[ $\xi \left(f_{1} ,f_{2} \right)$]  Nonlinear noise coefficient integrand.
\item[ $\nu_{k}$]  Normalized electric field attenuation coefficient.
\item[ $x_{k} \left(\zeta \right)$]  Normalized power complex attenuation coefficients.
\item[ $\zeta _{k} \left(\zeta \right)$] Normalized electric field phase shift.
\item[ $f_{\phi}$] Average phased-array bandwidth.
\item[ $f_{\phi _{k} }$] Phased-array bandwidth for the $k-$th fiber segment.
\item[ $\lambda _{k}$] Auxiliary multiplicative coefficients.
\item[ $B_0$] Optical bandwidth of the WDM signal.
\item[ $\Delta \nu _{{\rm res}}$] Resolution bandwidth.
\item[ $\sigma_k$] Normalized, chromatic dispersion-adjusted,  real attenuation coefficient for the $k-$th fiber segment.
\item[ $N_{\rm int}$] Number of periods of $\phi \left(\zeta \right)$ in the interval $\left[0,\zeta _{0} \right].$
\item[ $\Gamma$] Worst-case (real) effective nonlinear  coefficient.
\item[ $J(\mu,\zeta_0)$] Auxiliary integral, \\	$J(\mu,\zeta_0) \df \int_\mu^{\zeta_0}  N_s \phi(\zeta) \ln(\frac{\zeta_0}{\zeta})\eta(\zeta)  \, d\zeta$.
\item[ $R$ ] Region of integration in the hyperbolic $uv-$plane.
\item[$\epsilon_r$] Relative error.
\item[$I$] Various definite integrals. 
\item[$\Delta$] Step size of Simpson's quadrature.
\item[$N_{n} $] Number of integration nodes in a $\pi$ subinterval.
\item[$K_{N_s}(\delta)$] Auxiliary integral, \\ $K_{N_s}(\delta) \df \int_0^{\delta} \ln(\delta/\zeta) N_s \phi(\zeta) \, d\zeta .$
\item[$G_{\rm NLI} \left(f\right)$] Nonlinear noise psd.
\item[$\delta$] Small number in the vicinity of zero.
\end{IEEEdescription}

%\ifCLASSOPTIONcaptionsoff
%\newpage
%\fi




%\bibliographystyle{IEEEtran}


%\begin{thebibliography}{1}

%\end{thebibliography}

\bibliographystyle{ieeetr}
\bibliography{IEEEabrv,QSMFbib2} 

\end{document}